\documentclass[prb,aps,twocolumn,groupedaddress,floats,showpacs,final, fleqn]{revtex4-1}
\usepackage{graphicx}
\usepackage{amsmath}
\usepackage{dcolumn}
\usepackage{bm}
\usepackage{color}
\usepackage{mathtools}

\begin{document}
\definecolor{blue}{rgb}{0.3,0.3,0.9}
%%%%%%%%%%%%%%%%%%%%%%%%%%%%%%%%%%%%%% AUTHORS %%%%%%%%%%%%%%%%%%%%%%%%%

\author{Lode Pollet}
\affiliation{Department of Physics, Arnold Sommerfeld Center for Theoretical Physics and Center for NanoScience, University of Munich, Theresienstrasse 37, 80333 Munich, Germany}
%\affiliation{\color{red} Wilczek Quantum Center, Zhejiang University of Technology, Hangzhou 310014, China}
\author{Mikhail N. Kiselev}
\affiliation{The Abdus Salam International Centre for Theoretical Physics, Strada Costiera 11, I-34151 Trieste, Italy}
\author{Nikolay V. Prokof'ev}
\affiliation{Department of Physics, University of Massachusetts,
Amherst, MA 01003, USA}
\affiliation{Department of Physics, Arnold Sommerfeld Center for Theoretical Physics and Center for NanoScience, University of Munich, Theresienstrasse 37, 80333 Munich, Germany}
\affiliation{Department of Theoretical Physics, The Royal Institute of Technology, Stockholm SE-10691 Sweden}
\affiliation{National Research Center ``Kurchatov Institute,"
123182 Moscow, Russia}
\author{Boris V. Svistunov}
\affiliation{Department of Physics, University of Massachusetts,
Amherst, MA 01003, USA}
\affiliation{National Research Center ``Kurchatov Institute,"
123182 Moscow, Russia}
\affiliation{Wilczek Quantum Center, Zhejiang University of Technology, Hangzhou 310014, China}

%%%%%%%%%%%%%%%%%%%%%%%%%%%%%%%%%%%%%%%%%%%%%%%%%%%%%%%%%%%%%%%%%%%%%%%%%%%%%%

%\title{Applying Feynman diagrams  to classical bond models by representing strong-coupling expansions with Grassmann variables }
\title{Grassmannization of classical models}
\date{\today}
\begin{abstract}
Applying Feynman diagrammatics to non-fermionic strongly correlated models with local constraints might seem generically impossible for two separate reasons: (i) the necessity to have a Gaussian (non-interacting) limit on top of which the perturbative diagrammatic expansion is generated by Wick's theorem, and (ii) the Dyson's collapse argument implying that the expansion in powers of coupling constant is divergent. We show that for arbitrary classical lattice models both problems
can be solved/circumvented by reformulating the high-temperature expansion (more generally, any discrete representation of the model) in terms of Grassmann integrals. Discrete variables residing
on either links, plaquettes, or sites of the lattice are associated with the Grassmann
variables in such a way that the partition function (and correlations) of the original system and its  Grassmann-field counterpart are identical. The expansion of the latter around its Gaussian point generates Feynman diagrams. A proof-of-principle implementation is presented for the classical 2D Ising model. Our work paves the way for studying lattice gauge theories by treating bosonic and fermionic degrees of freedom on equal footing.

\end{abstract}

\maketitle

\section{Introduction}
Feynman diagrammatic technique is a powerful tool of statistical mechanics.  Among the hallmarks of the method are the ability to deal---both analytically and numerically---with the thermodynamic limit  rather than a finite-size cluster, the possibility of partial summations up to infinite order, and
the fully self-consistent formulation in terms of renormalized (dressed) quantities.
The latter properties allow one to go beyond the Taylor expansion in terms of the coupling constant
or any other parameter.

Advantages of the diagrammatic technique come at a price. The most serious issue is the
divergence of expansions in powers of the coupling constant for systems prone to
Dyson's collapse\cite{dyson1952divergence} ({\it i.e.}, pathological system behavior when the
coupling constant is rotated in the complex plane). For partial summation techniques to work,
the non-perturbed part of the theory has to be Gaussian 
(in terms of either real, or complex, or Grassmann variables) to ensure the validity of Wick's theorem.
These issues are often related: for example, Ising and XY models formulated in terms of original
spin variables do not suffer from Dyson's collapse but lack the Gaussian (non-interacting) limit,
while their classical (lattice) field counterparts with the well-defined Gaussian limit are subject
to Dyson's collapse. It would be a mistake, however, to think that meaningful
diagrammatic series are only possible for a very limited class of Hamiltonians, namely,
when the original system is that of interacting lattice fermions.
As already clearly explained by Samuel in a series of papers,\cite{Samuel_anticommuting1,Samuel_anticommuting2,Samuel_anticommuting3}
a broad class of classical spin and dimer models can be reformulated in terms of
familiar interacting fermions and studied with field-theoretical techniques.
Similarly, rather arbitrary quantum spin/boson lattice models can be rigorously
mapped onto fermionic field theories.\cite{PopFed,misha,PopovFedotovG}

As expected, grassmannian formulations of spin/link/boson models with local constraints
are generically strongly-coupled theories at low temperature, and even the most advanced
self-consistent treatments based on the lowest-order graphs are not supposed to provide
quantitatively (and often qualitatively) accurate answers.
Moreover, these theories may contain arbitrary multi-particle interaction vertexes, which 
further complicate the structure of the diagrammatic expansion.
One of the promising numerical techniques currently under development for strongly
correlated systems is diagrammatic Monte Carlo (DiagMC).
It is based on the stochastic evaluation of irreducible Feynman graphs up to some high order
and can be implemented in a number of ways, from perturbative expansions in powers
of the coupling constant to various self-consistent skeleton schemes based on fully renormalized
one- or two-body propagators. In such contexts as resonant fermions,\cite{VanHoucke2012}
frustrated magnetism,\cite{kulagin2013bdm, kulagin2013bdm2} and out-of-equilibrium impurity-like 
models\cite{Profumo2015, Cohen_out_of_eq_2015} the method was recently shown to be able to go significantly beyond the  state of the art. Also, significant progress has been
made in understanding superfluid properties of the Hubbard-type models.\cite{gukelberger2014pss,Deng2015,Gukelberger2015}
Notably, the infamous sign-problem
preventing conventional Monte Carlo methods from simulating fermionic system with sizes
large enough for reliable extrapolation to the thermodynamic limit, is absent as such in
DiagMC. Instead, the computational complexity is now linked to the number of diagrams
growing factorially with their order. Nevertheless, millions of diagrams can be accounted
for and the approach is flexible enough to deal with an arbitrary interaction Hamiltonian/action.

The current paradigm for generic lattice gauge models, as they occur in lattice-QCD as well
as in solid state and ultra-cold atomic physics, is to work with finite-size systems and
to treat link variables separately from the fermionic sector. More precisely, link
variables are simulated using classical Monte Carlo techniques (with local updates),
and fermions (quarks) are described by determinants. This approach suffers from a severe
sign-problem for finite density of fermions (non-zero chemical potential).\cite{QCDsignproblem,QCDsignproblem2}
If link variables are straightforwardly represented by bosonic fields, then the thermodynamic 
limit can be addressed within the diagrammatic approach that treats bosonic and fermionic degrees
of freedom on equal footing. However, in this formulation the bosonic fields pose a
fundamental problem, which manifests itself in a zero convergence radius. It is thus desirable
to have a generic scheme for replacing link variables with Grassmann fields to ensure that the
diagrammatic expansion has proper analytic properties around the Gaussian point.

In this paper, we introduce a general procedure of {\it grassmannization} for
classical lattice models. It is by no means a unique one, and in certain specific cases
more compact/simpler representations can be found. There is a strong connection to the anti-commuting variables approach introduced 
by S. Samuel,\cite{Samuel_anticommuting1, Samuel_anticommuting2, Samuel_anticommuting3} which can solve 
the 2D Ising model exactly (free fermion operators to solve the Ising model exactly were first found by Kaufman\cite{Kaufman1949} and 
refined by Schultz, Mattis and Lieb\cite{SchultzMattisLieb1964}) and provides a good starting point for field-theoretic studies of the 3D Ising model. 
For the latter system our approach amounts to an alternative but equally complicated field theory. 
Our prime goal is to build on these ideas and develop a scheme that is flexible enough to apply to a broader class of link models with arbitrary multi-bond
interactions and local constraints.

The idea of grassmannization is to represent the partition function of the model as a Grassmann
integral from the exponential of a Grassmann functional. The Feynman rules then emerge by
Taylor-expanding the non-Gaussian part of the exponential and applying Wick's theorem to the Gaussian averages. Paradigmatic lattice systems are link and plaquette 
models featuring discrete degrees of  freedom---integer numbers---residing on links (plaquettes) of square lattices and subject to certain
local constraints in terms of the allowed values of the sum of all link (plaquette) variables
adjacent to a given site (edge). It turns out that it is these constraints that require
special tricks involving multiple Grassmann variables for each value of each discrete variable.
Link models often emerge as high-temperature expansions of
lattice systems\cite{Oitmaa} in Ising, XY, O(3), etc. universality classes no matter whether the original
degrees of freedom are discrete or continuous (e.g., classical vector-field variables).
Link models may also emerge as dual (low-temperature) expansions, and specific examples
are provided by the 2D Ising model\cite{mccoy1973} and the 3D $|\psi|^4$ model (the latter case leads to the
so-called J-current model with long-range interactions). Similarly, plaquette models emerge as a
high-temperature expansion of lattice gauge theories, but sometimes they represent the dual
(low-temperature) expansion, as in the case of the 3D Ising model. Finally, it is worth mentioning
how the models with the same general structure are generated by strong-coupling expansions
in lattice-QCD.\cite{Wilsonloop}

The paper is structured as follows. In Sec.~\ref{sec:II} we explain how a partition function
of a discrete link model can be written as a Grassmann integral. The equivalence between the two
formulations is readily proved through term-by-term comparison. Standard properties of Grassmann
variables then immediately allow one to express the Grassmann weight in the exponential form in order to
define the field-theory. In Sec.~\ref{sec:III} we discuss generalizations of the proposed
grassmannization scheme. We start by describing the procedure for a broad class of plaquette models.
Next we show a simple way to introduce Grassmann variables for non-local link models with pairwise
interactions between the link variables. The construction is further simplified when constraints
are replaced with statistical penalties for certain configurations of link (plaquette) variables.
We conclude this section with defining the meaning of the term ``order of expansion" for the
resulting field theory. In Sec.~\ref{sec:IV} we deliberately choose the most general
grassmannization scheme for the 2D Ising model to illustrate and test how our construction
works in practice. We stress that our goal is not to solve the 2D Ising model
exactly\cite{Onsager44, mccoy1973} or determine a series expansion
for it\cite{Oitmaa} but to develop a general framework---including numeric component---for applying Grassmann variables
to link and plaquette models and show that its evaluation can be done realistically. After determining all field-theoretic parameters, characterizing various
interaction terms and source operators for calculating correlation functions (with and without magnetic
field), and explaining Feynman rules for constructing the perturbative expansion, we proceed with
the description of algorithms to compute them  (Monte Carlo and deterministic)
in Sec.~\ref{sec:V}. Results are presented and discussed in Sec.~\ref{sec:VI}.
By comparing with the exact solution we show that the critical exponent
$\gamma$ for magnetic susceptibility could be determined with an accuracy of about $5$\%, while
the critical point could be located with sub-percent accuracy. In Sec.~\ref{sec:VII} we discuss the
implementation of the self-consistent skeleton technique within the so-called $G^2W$-expansion\cite{Heidin}
which computes irreducible (skeleton) diagrams for the self-energy and ``polarization" function
and uses them in the Dyson equations in order to find the renormalized propagators and screened interactions.
We also present results that emerge when this technique is based on several low-order diagrams.
We briefly comment in Sec.~\ref{sec:VIII} that both bare-series and the $G^2W$-expansion methods readily
solve the 1D Ising model exactly. We conclude with prospects for future work in Sec.~\ref{sec:IX}.

%%%%%%%%%%%%%%%%%%%%%%%%%%%%%%%%%%%%%%%%%
%\section{Deriving the diagrammatic technique}
\section{Grassmannization of local link models}
\label{sec:II}

%%%%%%%%%%%%
\subsection{Local link models}

For the purposes of this article,  we mean by a link model a classical statistical model
with states labeled by a set of {\it discrete} variables $\{ \alpha_b \}$ residing on
links (bonds) of a certain lattice. In addition, we require that the ground state
is unique. Without loss of generality, it can be chosen to be the state with $\alpha_b=0$
on each link $b$.

We further narrow the class of link models---to which we will refer to as {\it local} link models---by the requirement that the statistical weight of a state factors into a product of link and site weights
(to be referred to as link and site factors, respectively). A link factor, $f_b$, is a function
of the corresponding link variable, $f_b\equiv f(\alpha_b)$.  The site factor, $g_j$, is a function that
depends on all variables residing on links attached to the site $j$, denoted as $\{ \alpha_b \}_j$. Then, $g_j \equiv g(\{ \alpha_b \}_j)$.
Solely for the purpose of avoiding heavy notations, we consider translational invariance when  $f(\alpha_b) \equiv f_b$ is the same function  on all links and $g_j$ site independent, $g_j \equiv g$.
Given that only the relative weights of the states matter,
we set $f(0)=1$ and $g(0_j)=1$, where $0_j$ stands for the $\{ \alpha_b=0 \}_j$ set.

The site factors play the key role in link models. They describe interactions between
(otherwise independent) link degrees of freedom. In particular, this interaction can take the extreme form
of a {\it constraint} on the allowed physical configurations of $\{ \alpha_b \}_j$ ({\it e.g.,} the zero-divergency constraint in J-current models,\cite{Villain}
or the even-number constraint in the high-temperature expansion
of $Z_2$ models), in which case $g_j(\{ \alpha_b \}_j)$ is identically {\it zero} for each non-physical state
of $\{ \alpha_b \}_j$.

%%%%%%%%%%%%%%%%%%%%%%%%%%%%%%
\subsection{Grassmannization}

%%%%%%%%%%%%%%%%%%%%%%%%%%%%%%%%%%%%%%%%%
\begin{figure}[tbp]
\centering
\includegraphics[width=0.18\columnwidth, angle=-90]{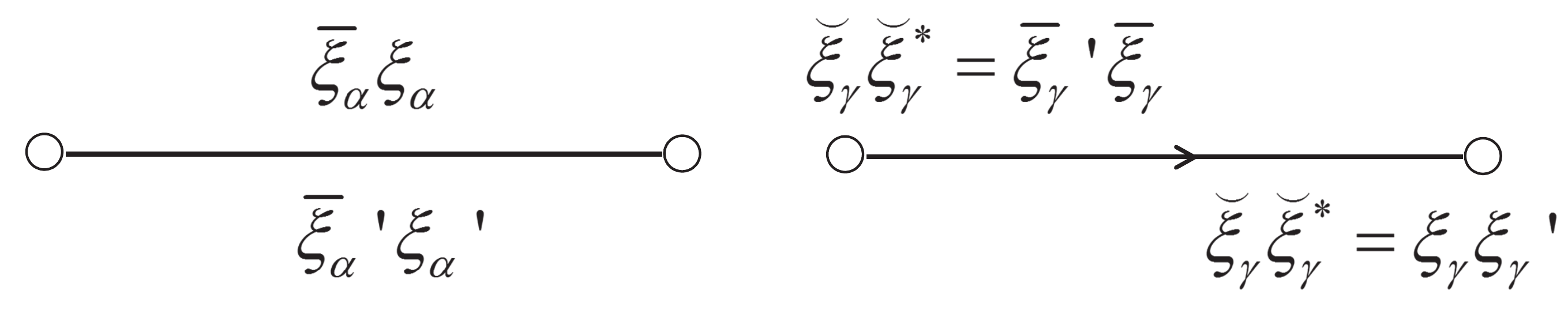}
\caption{Assignment of Grassmann fields for link (left) and site (right) factors.
Upon integration, the labels of the Grassmann variables must be equal in order to
connect variables from all factors (see text).
}
\label{fig:fields}
\end{figure}
%%%%%%%%%%%%%%%%%%%%%%%%%%%%%%%%%%%%%%%%%

For each label $\alpha \neq 0$ of the link $b$, introduce four Grassmann variables: $\xi_{\alpha,b}$, $\xi_{\alpha,b}'$, $\bar{\xi}_{\alpha,b}$, and $\bar{\xi}_{\alpha,b}'$. For a textbook
introduction to Grassmann variables, we refer to Ref.~\onlinecite{negele1988quantum}.
For $\alpha = 0$ we assume that
$\xi_{0,b}=\xi_{0,b}'=\bar{\xi}_{0,b}=\bar{\xi}_{0,b}'=1$.
In terms of these variables, define the Grassmann weight---a product of link, $A_b$, and site, $B_j$, factors such that tracing over all degrees of freedom yields the partition function $Z = {\rm Tr} \prod A_b \prod B_j$---by the following rules,
\begin{eqnarray}
A_b & = & \exp \left\{ \sum_{\alpha \neq 0}  \left[
\frac{\bar{\xi}_{\alpha, b}' \xi_{\alpha, b}'}{\sqrt{f(\alpha)}} +
\frac{\bar{\xi}_{\alpha, b} \xi_{\alpha, b}}{\sqrt{ f(\alpha)}} \,   \right]\right\} \nonumber \\
   & = & \prod_{\alpha \neq 0}
\, \exp \left\{ \frac{\bar{\xi}_{\alpha, b}' \xi_{\alpha, b}'}{\sqrt{f(\alpha)}} +
             \frac{\bar{\xi}_{\alpha, b} \xi_{\alpha, b}}{\sqrt{ f(\alpha)}}  \right\} \, ,
\label{link}
\end{eqnarray}
\begin{eqnarray}
B_j\, =\,  \sum_{ \{ \alpha_b \}_j} g(\{ \alpha_b \}_j) \prod_{b \in \{ b \}_j} \breve{\xi}_{\alpha_{b}, b}\,  \breve{\xi}_{\alpha_{b}, b}^*\nonumber \\
 \, =\, 1  \, + \sum_{ \{ \alpha_b \}_j \neq 0_j}  g(\{ \alpha_b \}_j) \prod_{b \in \{ b \}_j} \breve{\xi}_{\alpha_{b}, b}\,  \breve{\xi}_{\alpha_{b}, b}^* \, .
\label{site}
\end{eqnarray}
Here $\{ b \}_j$ stands for the set of all links incident to the site $j$, and variables
$\breve{\xi}_{\alpha_{b}, b}$ and  $\breve{\xi}_{\alpha_{b}, b}^*$ are defined differently
for different links. We first introduce the notion of direction (on each link)
so that one of the two link ends becomes ``incoming" and its counterpart ``outgoing"
(with respect to the site adjacent to the end). Next, we assign (see Fig.~\ref{fig:fields} for an illustration)
\begin{equation}
\begin{array}{*{5}{l}}
\breve{\xi}_{\alpha_b, b} =\xi_{\alpha_b, b}\, , \;\; \breve{\xi}_{\alpha_b, b}^*= \xi_{\alpha_b, b}' \; \;   \mbox{(for incoming end)} ,\\
\breve{\xi}_{\alpha_b, b} =\bar{\xi}_{\alpha_b, b}'\, , \;\; \breve{\xi}_{\alpha_b, b}^*= \bar{\xi}_{\alpha_b, b} \; \;  \mbox{(for outgoing end) .}
\end{array}
\label{convention}
\end{equation}

The claim is that the Grassmann integral of the weight over all variables reproduces the partition function of the original link model. For a link $b$ to yield a non-zero contribution to the integral the link labels in (\ref{site}) for the sites of the incoming ($j=1$) and outgoing ($j=2$) ends of the link should match each other:
$\alpha_1=\alpha_2$. Indeed, at $\alpha_1\neq \alpha_2$, it is not possible to find
an appropriate term in the expansion of the link exponential (\ref{link})
such that---upon multiplying by the site factors $\breve{\xi}_{\alpha_1, b} \, \breve{\xi}_{\alpha_1, b}^* $ and $\breve{\xi}_{\alpha_2, b} \, \breve{\xi}_{\alpha_2, b}^*$---all powers of the Grassmann variables $\xi_{\alpha_1,b}$, $\xi_{\alpha_1, b}'$, $\bar{\xi}_{\alpha_1, b}$,  $\bar{\xi}_{\alpha_1, b}'$, $\xi_{\alpha_2, b}$, $\xi_{\alpha_2, b}'$, $\bar{\xi}_{\alpha_2, b}$,  $\bar{\xi}_{\alpha_2, b}'$ are exactly equal to 1 to ensure that the Grassmann integral is non-zero.
For $\alpha_1=\alpha_2\equiv \alpha$,
we need to consider two cases: $\alpha =0$ and $\alpha \neq 0$.
In the first case, the non-zero contribution to the integral comes from the product of second terms in the expansion of the link exponentials (\ref{link}):
\begin{eqnarray}
&\!& \!\!\!\!\!\!\!\!
\prod_{\gamma \neq 0}  \int  \mathcal{D} [ \bar{\xi}' \xi' \bar{\xi} \xi ]_{\gamma }
\, \exp \left\{ \frac{\bar{\xi}_{\gamma}' \xi_{\gamma}'}{\sqrt{f(\gamma)}} +
             \frac{\bar{\xi}_{\gamma} \xi_{\gamma}}{\sqrt{ f(\gamma)}}  \right\}
\nonumber \\
&=&\prod_{\gamma \neq 0}  \int  \mathcal{D} [ \bar{\xi}' \xi' \bar{\xi} \xi ]_{\gamma }
\left[ 1 + \frac{\bar{\xi}_{\gamma}' \xi_{\gamma}'}{\sqrt{f(\gamma)}} \right] \,
\left[ 1 + \frac{\bar{\xi}_{\gamma} \xi_{\gamma}}{\sqrt{ f(\gamma)}} \right]  \nonumber \\
 &=&  \prod_{\gamma \neq 0} \frac{1}{f(\gamma)} \equiv \frac{1}{f_*} \;,
\label{f_star}
\end{eqnarray}
where we defined $f_*$ in the last step.
%
%\begin{equation}
%\prod_{\gamma } {1\over f(\gamma)}\,   \bar{\xi}_{\gamma, b}' \, \xi_{\gamma, b}' \,  %\bar{\xi}_{\gamma, b} \, \xi_{\gamma, b}\, , \quad (\mbox{at}~~\alpha =0)  .
%\label{groundstate}
%\end{equation}
%Upon integration, it contributes a (link-independent) factor $1/f_*$, where
%\begin{equation}
%f_* = \prod_{\gamma } f(\gamma) \, .
%\label{f_star}
%\end{equation}
In the second case, the two end sites contribute the factor $\xi_{\alpha, b}\,  \xi_{\alpha, b}'\,  \bar{\xi}_{\alpha, b}' \, \bar{\xi}_{\alpha, b}=\bar{\xi}_{\alpha, b}' \, \xi_{\alpha, b}' \, \bar{\xi}_{\alpha, b} \, \xi_{\alpha, b}$. Now we have to consider the first term in the expansion
of the link exponential for state $\alpha$, while for other variables the calculation is repeated
as in (\ref{f_star})
\begin{eqnarray}
&\!& \!\!\!\!\!\!\!
\prod_{\gamma \neq 0}        \int \mathcal{D} [ \bar{\xi}' \xi' \bar{\xi} \xi ]_{\gamma }
\bar{\xi}'_{\alpha} \bar{\xi}_{\alpha}
\left[ 1 + \frac{\bar{\xi}_{\gamma}' \xi_{\gamma}'}{\sqrt{f(\gamma)}} \right] \,
\left[ 1 + \frac{\bar{\xi}_{\gamma} \xi_{\gamma}}{\sqrt{ f(\gamma)}} \right]
\xi_{\alpha} \xi'_{\alpha} \nonumber \\
 & = &  \prod_{\gamma \neq 0, \alpha}  \frac{1}{f(\gamma)} = \frac{f(\alpha)}{f_*}.
\label{non_groundstate}
\end{eqnarray}
We see that, apart from the irrelevant global factor $\prod_b 1/f_*$,
we reproduce the configuration space and weight factors of the original link model.

%%%%%%%%%%%%%%%%%%%%%%%%%%%%%%%%%%%%%
\subsection{Field-theoretical formulation}

To generate the Feynman diagrammatic expansion, we need to represent the Grassmann weight factor in the exponential form. The link factors (\ref{link}) have the form of Gaussian exponentials
already. Hence, it is only the site factors that need to be rewritten identically as
\begin{equation}
B_j\, =\, \exp \left[  \sum_{ \{ \alpha_b \}_j } \lambda(\{ \alpha_b \}_j) \prod_{b \in \{ b \}_j} \breve{\xi}_{\alpha_b, b} \, \breve{\xi}_{\alpha_b, b}^*
\right] \, .
\label{site2}
\end{equation}
The constants $ \lambda(\{ \alpha_b \}_j)$ are readily related to the site
factors $g(\{ \alpha_b \}_j)$ by simple algebraic equations obtained by expanding
the exponential and equating sums of similar terms to their counterparts in the r.h.s. of  Eq.~(\ref{site}).

By expanding the non-Gaussian part of the exponential (\ref{site2}) and applying Wick's theorem,
we arrive at Feynman rules for the diagrammatic series.
The reader should avoid confusion by thinking that an expansion of the exponential (\ref{site2})
takes us back to Eq.~(\ref{site}). Recall that connected Feynman diagrams are formulated
for the free energy density, not the partition function, and summation over
all lattice sites is done for a given set of interaction vertexes in the graph,
as opposite to the summation over all vertex types for a given set of lattice points.
Therefore, the ``coupling constants" in Feynman diagrams are $\lambda$'s, not $g$'s.

%%%%%%%%%%%%%%%%%%%%%%%%%%%%%%%%%%%%%%%%%%%%%%%%%%%%%
\subsection{Absorbing link factors into site factors}

The separation of the weight factors into link and site ones is
merely a convention. Indeed, each link factor can be ascribed
to one of the two site factors at its ends. This leads to a
slightly different Grassmannization protocol.
% in which there are no terms inversely proportional to $f_\alpha$.
This trick may prove convenient for generalization to non-local models
considered below.

%%%%%%%%%%%%%%%%%%%%%%%%%
\section{Generalizations}
\label{sec:III}

%%%%%%%%%%%%%%%%%%%%%%%%%%%%%
\subsection{Plaquette models}

A plaquette model can be viewed as a certain generalization of the local link model.
States (configurations) of a plaquette model are indexed by a set of discrete labels
residing on (oriented) plaquettes of a hyper-cubic lattice.
The plaquette label $\alpha$ takes on either a finite or countably infinite number of values.
The statistical weight of each state factors into a product of plaquette and edge weights (to be referred to as plaquette and edge factors, respectively). A plaquette  factor, $f$, is a  function of the corresponding plaquette  variable,  $f\equiv f(\alpha)$. An edge factor, $g$, is a function which depends on the labels of all plaquettes sharing this edge (this set of labels will be denoted as
$\{ \alpha_p \}_j$ for the edge $j$); it encodes, if necessary, constraints on the allowed
sets of $\{ \alpha_p \}_j$.

Without loss of generality (up to a global normalization factor), we  identify the
``ground state" as $\alpha_p = 0$ for all plaquettes, and set $f(0)=1$.
The orientation of the plaquette (for some models it is merely a matter of convenience)
is enforced by an ordered enumeration of sites at its boundary. For a plaquette $p$, the vertex label $\nu \equiv \nu_p = 0,\, 1,\, 2,\, 3$ enumerates four vertices in such a way that $\nu \pm 1$ modulo 4 stands for the  next/previous vertex with respect to the vertex $\nu$ in the clockwise direction.

For each  state $\alpha \neq 0$ of the plaquette $p$, we introduce eight Grassmann variables: $\xi_{\alpha,p, \nu_p}$, $\bar{\xi}_{\alpha,p, \nu_p}, \nu_p = 0,1,2,3$. 
As before, for $\alpha=0$ the variables $\xi$ and $\bar{\xi}$ are not Grassmannian, $\xi_{0, p, \nu } =0$, $\bar{\xi}_{0,p,\nu }=1$.
The corresponding plaquette weight in the Grassmann partition function reads
\begin{equation}
A_p = \exp \left\{  \sum_{\alpha \neq 0}\,   [-f(\alpha)]^{-1/4} \sum_{\nu_p=0}^{3}\,   \bar{\xi}_{\alpha,p, \nu_p}\, \xi_{\alpha,p, \nu_p} \right\}\, .
\label{plaquette}
\end{equation}
Note a close analogy with Eq.~(\ref{link}).
%, especially if an alternate form of (\ref{link}) is used in which $1/f(\alpha)$ factor is split ``democratically" between the pairs of primed and non-primed  
% variables so that each pair comes with the factor $[f(\alpha)]^{-1/2}$. The necessity of taking $f(\alpha)$ with the minus sign will become clear later.
Site weights Eq.~(\ref{site}) are now replaced with edge weights $B_j$.
Using the notation $\{ p \}_j$ for the set of all plaquettes sharing the
edge $j$, and $0_j$ for the state when all plaquettes in this set have
$\alpha_p=0$, we write
\begin{equation}
B_j\, =\, 1 + \sum_{ \{ \alpha_p \}_j \neq 0_j} g(\{ \alpha_p \}_j) \prod_{p \in \{ p \}_j}  \xi_{\alpha,p, (\nu_p^{(j)}+1)} \, \bar{\xi}_{\alpha,p, \nu_p^{(j)}}\, ,
\label{edge}
\end{equation}
where $\nu_p^{(j)}$ is the site enumeration index within the plaquette $p$, with respect to which the edge $j$ is outgoing. [Accordingly, the edge $j$ is incoming with respect to site $(\nu_p^{(j)}+1)$.] In what follows, we will associate $\nu_p^{(j)}$ not only with the site, but also with the corresponding edge.

The proof that the classical and Grassmannian partition functions are identical (up to a global factor) is similar to the one for the link model after we notice that a non-zero contribution from
plaquette $p$ is possible only if the same plaquette label $\alpha_p$ is used in all edge weights.
The $\alpha =0$ contribution comes from the term
\begin{equation}
-\prod_{\gamma} {1\over f(\gamma)}\,   \prod_{\nu_p=0}^{3}\,  \bar{\xi}_{\gamma, p, \nu_p} \, \xi_{\gamma, p, \nu_p} \qquad (\mbox{at}~~\alpha =0) 
\label{groundstate_p}
\end{equation}
in the expansion of the exponential (\ref{plaquette}). It contributes a factor $1/q_*$,  where
\begin{equation}
q_* = \prod_{\gamma } (-1) f(\gamma) \, .
\label{f_star_p}
\end{equation}
The $\alpha \neq 0$ contribution comes from the plaquette term
\begin{equation}
\prod_{\gamma \neq \alpha} {1\over f(\gamma)}\,   \prod_{\nu_p=0}^{3}\,  \bar{\xi}_{\gamma, p, \nu_p} \, \xi_{\gamma, p, \nu_p}  \qquad (\mbox{at}~~\alpha \neq 0)
\label{non_groundstate_p}
\end{equation}
multiplied by the product $\prod_{\nu_p=0}^{3}\,  \bar{\xi}_{\alpha, p, \nu_p} \, \xi_{\alpha, p, \nu_p}$ originating from the boundary edge terms $ \xi_{\alpha,p, (\nu_p^{(j)}+1)} \, \bar{\xi}_{\alpha,p, \nu_p^{(j)}}$. Because of the Grassmann anticommutation rules, this four-edge factor yields an additional minus sign, explaining the use of the negative sign
in front of $f(\alpha)$ in Eq.~(\ref{plaquette}).
Upon Grassmann integration, the contribution to the partition function of the resulting term equals to $f(\alpha)/q_*$.

Feynman diagrammatics for the plaquette model is obtained by following the same basic steps as for
the link models. The Gaussian part is given by Eq.~(\ref{plaquette}) with four pairs of Grassmann
fields for every non-zero plaquette state.
% each of which connects the edge $\nu$ with the edge $\nu+1$. The quartet of propagators enters the
% diagram as an integral piece carrying the factor $f(\alpha)$, where $\alpha$ is the ``flavor" of the
% propagators. Graphically, the quartet can be represented as a four-pole object, each pole ending at
% the corresponding edge of the plaquette.
The interaction part of the Grassmann action is contained in edge weights (\ref{edge})
after they are written in an exponential form
\begin{equation}
B_j\, =\, \exp \left[  \sum_{ \{ \alpha_p \}_j} \lambda(\{ \alpha_p \}_j) \prod_{p \in \{ p \}_j}  \xi_{\alpha,p, (\nu_p^{(j)}+1)} \, \bar{\xi}_{\alpha,p, \nu_p^{(j)}} \right] \, ,
\label{edge2}
\end{equation}
with the constants $ \lambda(\{ \alpha_b \}_j)$ unambiguously related to the edge factors $g(\{ \alpha_b \}_j)$.

%%%%%%%%%%%%%%%%%%%%%%%%%%%%%%%%%%%%%%%%%%%%%%%%%%%%%%%%%%%%%%%%%%%%%%
\subsection{Unconstrained discrete models with pair-wise interaction}

The hallmark of the considered link (plaquette) models is the non-trivial interaction introduced via site (edge) factors. It is due to this type of interaction---and, in particular, its extreme form of a constraint on allowed combinations of discrete variables---that we had to introduce multiple
Grassmann variables for each state of the link (plaquette). The situation simplifies dramatically
if we are dealing with unconstrained discrete degrees of freedom with pair interactions between them.

Consider a link model defined by the statistical weight
\begin{equation}
W(\{ \alpha_b \}) = \prod_{b_1 , b_2}
                    F(\, \alpha_{b_1}, b_1; \, \alpha_{b_2}, b_2 ) \;,
\label{Wlink22}
\end{equation}
based on products of two-link factors. Without loss of generality, these factors can be
cast into the exponential form
\begin{equation}
W(\{ \alpha_b \}) = \prod_{b_1 , b_2}\,
                    e^{-(1/2) \eta_{\, \alpha_{b_1}, b_1;\, \alpha_{b_2}, b_2}} \;,
\label{Wlink2}
\end{equation}
We assume that all factors in the product are bounded and properties of the $\eta$-matrix are well-conditioned.
%
% It would be wrong to think that an example is
% provided by the dual representation of the 3D XY universality
% class---its low-temperature expansion is given by a J-current model with pairwise
% Coulomb interaction between the link currents.
% because it also has a zero divergence constraint
%
Grassmannization of this model can be done by taking advantage of properties of Gaussian integrals
that allow one to express (\ref{Wlink2}) identically (up to normalization) as
\begin{equation}
W(\{ \alpha_b \}) = \int {\cal D} X  \prod_{b} \, e^{iX_{\,\alpha_b, b}}
\, W_G( \{ X_{\, \alpha_b, b} \} )  \;.
\label{Wlink3}
\end{equation}
Here $\{ X_{\, \alpha_b, b} \}$ is a collection of auxiliary real continuous variables.
For briefness, we do not show explicitly the Gaussian weight $W_G$ that is
uniquely defined by the values of all pairwise averages performed with this weight
\begin{equation}
\eta_{\, \alpha_{b_1}, b_1; \,\alpha_{b_2}, b_2} \, =\, \langle X_{ \alpha_{b_1}, b_1}
 X_{\alpha_{b_2}, b_2 } \rangle \, .
\label{expon1}
\end{equation}

%What we achieve under the sign of Gaussian integration over $X$ variables is a link model that contains only single-link factors
What we achieve for a fixed set of $X$ variables is a link model that contains only single-link factors
\begin{equation}
\forall b: \qquad f_b(\alpha_b) = \, e^{iX_{\alpha_b,b }}.
\label{expon}
\end{equation}
For models with site constraints, link factors can be attributed to site factors 
at the incoming (or outgoing) ends with subsequent Grassmannization of the latter
as discussed above. For unconstrained models, Grassmannization is accomplished by replacing sums over link variables with
\begin{equation}
\sum_{\alpha_b} f_b(\alpha_b) \, \to \,
{\cal W}^{(G)}_b  \, =\,  \exp \left[  \bar{\xi}_b \xi_b \left( \sum_{\alpha_b} \, e^{iX_{\, \alpha_b, b}} \right)  \right] \;.
\label{weight_GL}
\end{equation}
Note that here Grassmann variables have nothing to do with the discrete index $\alpha_b$,
in contrast with previous considerations. The resulting formulation contains both
Grassmann and real-number integrations.

Clearly, all considerations can be repeated identically (up to a trivial change in notations)
for a model based on discrete variables  $\alpha_s$ residing on lattice sites when
the configuration weight is given by
\begin{equation}
W(\{ \alpha_s \}) = \prod_{s_1 , s_2}\,
                    e^{-(1/2) \eta_{\, \alpha_{s_1}, s_1;\, \alpha_{s_2}, s_2}} \; .
\label{Q_1_Q_2}
\end{equation}

%%%%%%%%%%%%%%%%%%%%%%%%%%%%%%%%
\subsection{Order of expansion}
\label{subsec:D}

The notion of the order of expansion is absolutely central for practical applications
when diagrammatic series are truncated. Normally, it is defined as an integer non-negative power of a certain dimensionless parameter $\zeta$ playing the role of a generalized coupling constant,
such that the diagrammatic expansion corresponds to a Taylor expansion in $\zeta$ about the point $\zeta=0$. Without loss of generality,
we can always select $\zeta$ (by an appropriate rescaling) in such a way that the physical value of $\zeta$ is 1. This is especially convenient in cases
when there is more than one interaction vertex, and ascribing different powers of
$\zeta$ to them results in (re-)grouping of different terms in the series.
A reasonable guiding principle behind such a (re-)grouping is the requirement to end up with
Taylor series having finite convergence radius around $\zeta=0$.
The latter is guaranteed if the theory is analytic in $\zeta$ at the origin; the necessary condition for this to be true is the absence of Dyson's collapse when changing
the sign (more generally, the phase) of $\zeta$.

As an illustration, consider the theory (\ref{Wlink22})-(\ref{Wlink2}) and its
Grassmann counterpart (\ref{weight_GL}). Introduce the $\zeta$-dependence by the replacement
\begin{equation}
 e^{iX_{\, \alpha_b, b } } \, \to \,  e^{i\zeta X_{\, \alpha_b, b}} .
\label{zeta}
\end{equation}
In terms of the original theory, the replacement (\ref{zeta})
means $\eta \to \zeta^2 \eta$, for all $\eta$'s in Eq.~(\ref{Wlink2}).
If amplitudes of all $\eta$ values in (\ref{Wlink2}) are bounded,
we expect that such a dependence on $\zeta$ is analytic not only for a finite system,
but also in the thermodynamic limit at finite temperature.
In the Grassmann action (\ref{weight_GL}), the expansion of the exponential $e^{i\zeta X_{s, \alpha_b}}$ in powers of $\zeta$ generates an infinite series of interaction vertexes (the zeroth-order term defines the harmonic action):
\begin{equation}
\bar{\xi}_b \xi_b \sum_{\alpha_b } \left( i \zeta  X_{\, \alpha_b, b} - {1\over 2} \zeta^2
 X_{\, \alpha_b, b}^2 -  {i\over 3!} \zeta^3 X_{\, \alpha_b, b}^3 + \ldots  \right).
\label{coupling}
\end{equation}
Higher-order vertexes in $X$ come with a higher power of $\zeta$ and this sets unambiguously the rules for defining the diagram order.

%%%%%%%%%%%%%%%%%%%%%%%%%%%
\section{Illustration for the 2D Ising model}
\label{sec:IV}

%%%%%%%%%%%%%%%%%%%%%%%%%%%%%%%%%%%%%%%%%
\begin{figure}[tbp]
\centering
 \includegraphics[width=1.0\columnwidth]{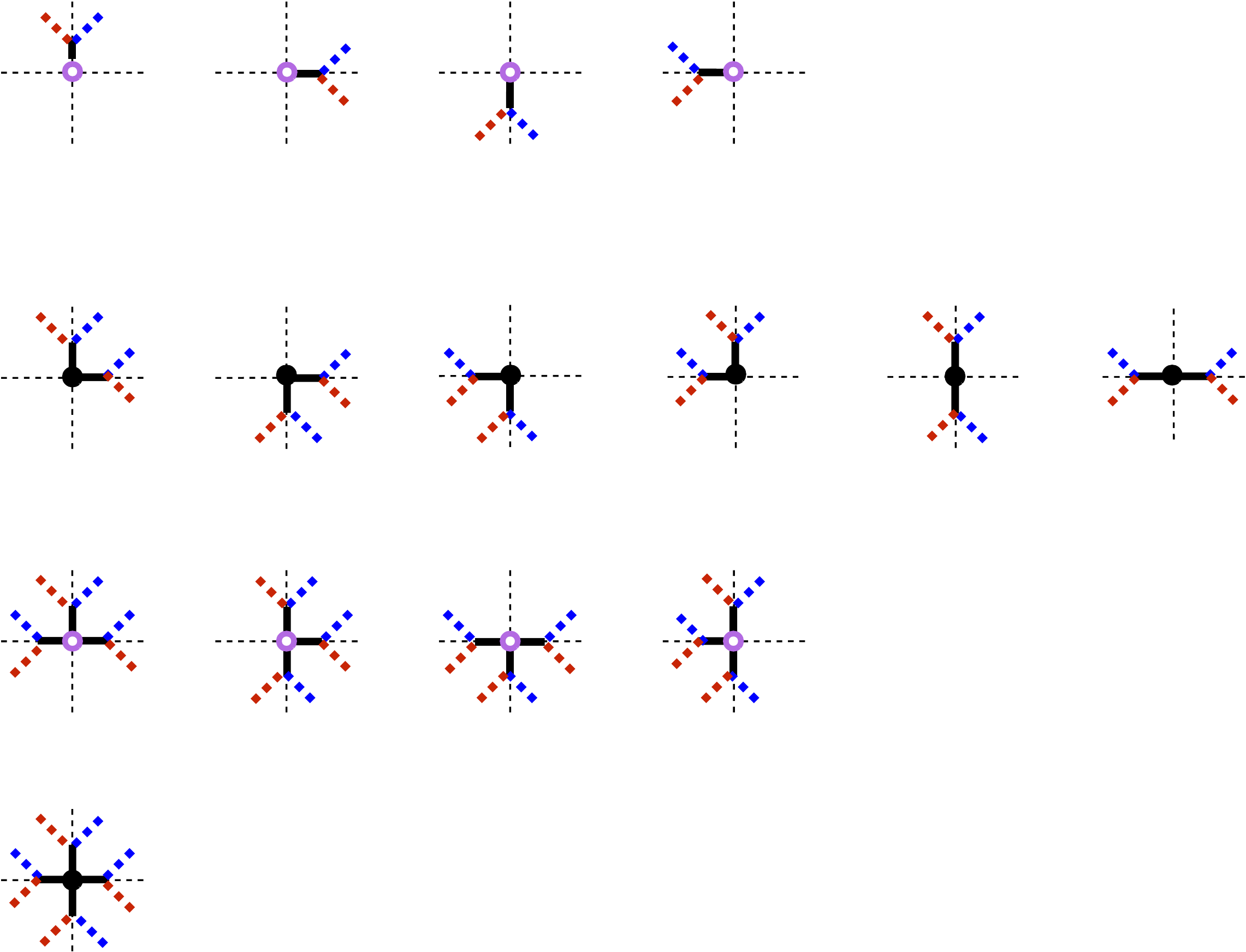}
\caption{(Color online) Four classes of generic pre-diagrams for link models on a square lattice. The elements in the first and the third row can only occur at the end points of the spin correlator (indicated by the open circle), the elements in the second and fourth row are the generic basic vertexes of the theory ascribed to the sites of the underlying lattice. There are hence 4 $V_1$ vertexes with 1 leg (first row, $U, R, D,$ and $L$), 6 $V_2$ vertexes with 2 legs (second row, $RU, RD, LD, LU, UD$ and $LR$), 4 $V_3$ vertexes with 3 legs (third row, $LUR, URD, LDR,$ and $DLU$), and 1 $V_4$ vertex with 4 legs (fourth row, $RULD$).  Connected to the legs of these vertexes are pairs of bi-Grassmann fields (thick dash lines (blue and red)) that reside on the links of the underlying 2D lattice. Thin dashed lines (showing lattice links adjacent to the site of the vertex) are to guide the eye and have no other meaning than showing the underlying 2D lattice.  The generalization to other dimensions is straightforward.} \label{fig:prediagrams}
\end{figure}
%%%%%%%%%%%%%%%%%%%%%%%%%%%%%%%%%%%%%%%%%

 %%%%%%%%%%%%%%%%%%%%%%%%%%%%%%%%%%%%%%%%%
\begin{figure}[tbp]
\centering
 \includegraphics[trim = 0 0mm 0mm 0mm, clip, width=1.0\columnwidth]{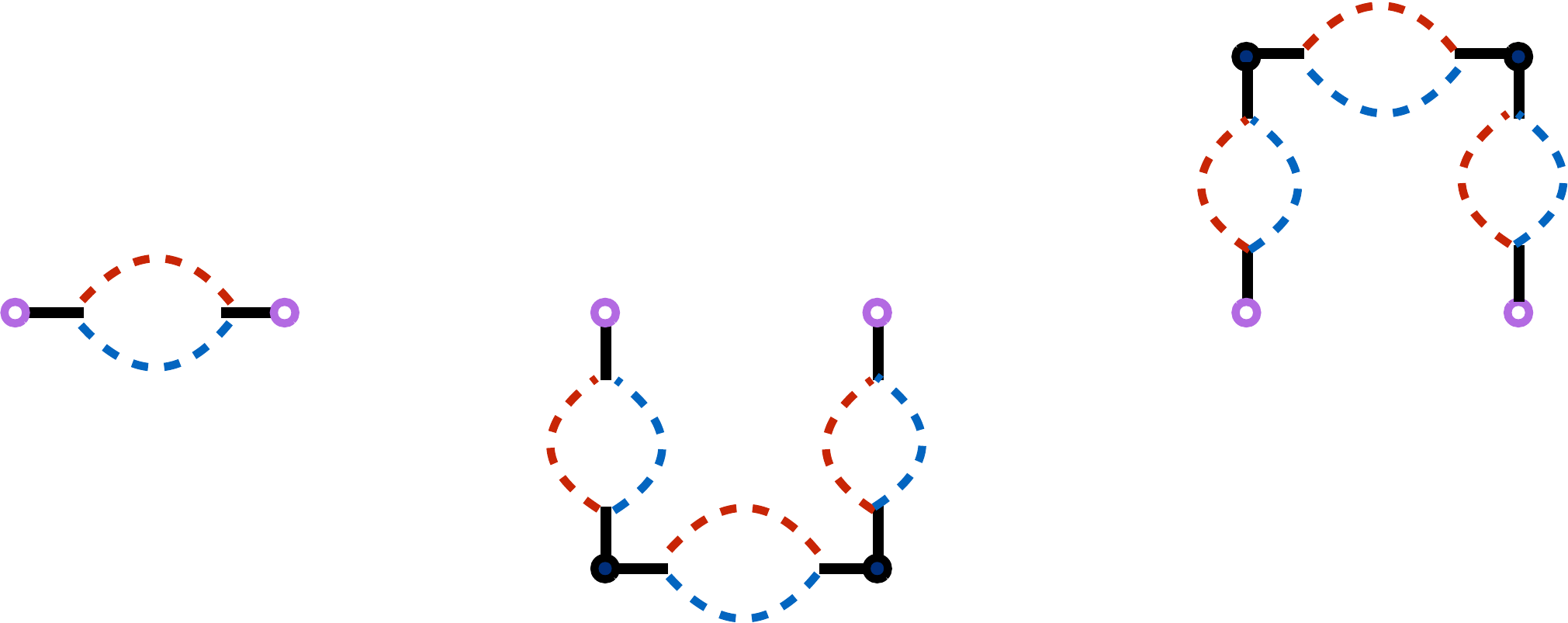}
 %trim option's parameter order: left bottom right top
\caption{(Color online) The first- and third-order diagrams for $\rho_{(1,0)}$ (at $h=0$) based on expanding (\ref{sigma_sigma_correlator_1}). The contribution of these diagrams is $\zeta+ 2\zeta^3$.}
\label{fig:rho_generic_a}
\end{figure}
%%%%%%%%%%%%%%%%%%%%%%%%%%%%%%%%%%%%%%%%%

 %%%%%%%%%%%%%%%%%%%%%%%%%%%%%%%%%%%%%%%%%
\begin{figure}[tbp]
\centering
 \includegraphics[trim = 0 0mm 0mm 0, clip, width=1.0\columnwidth]{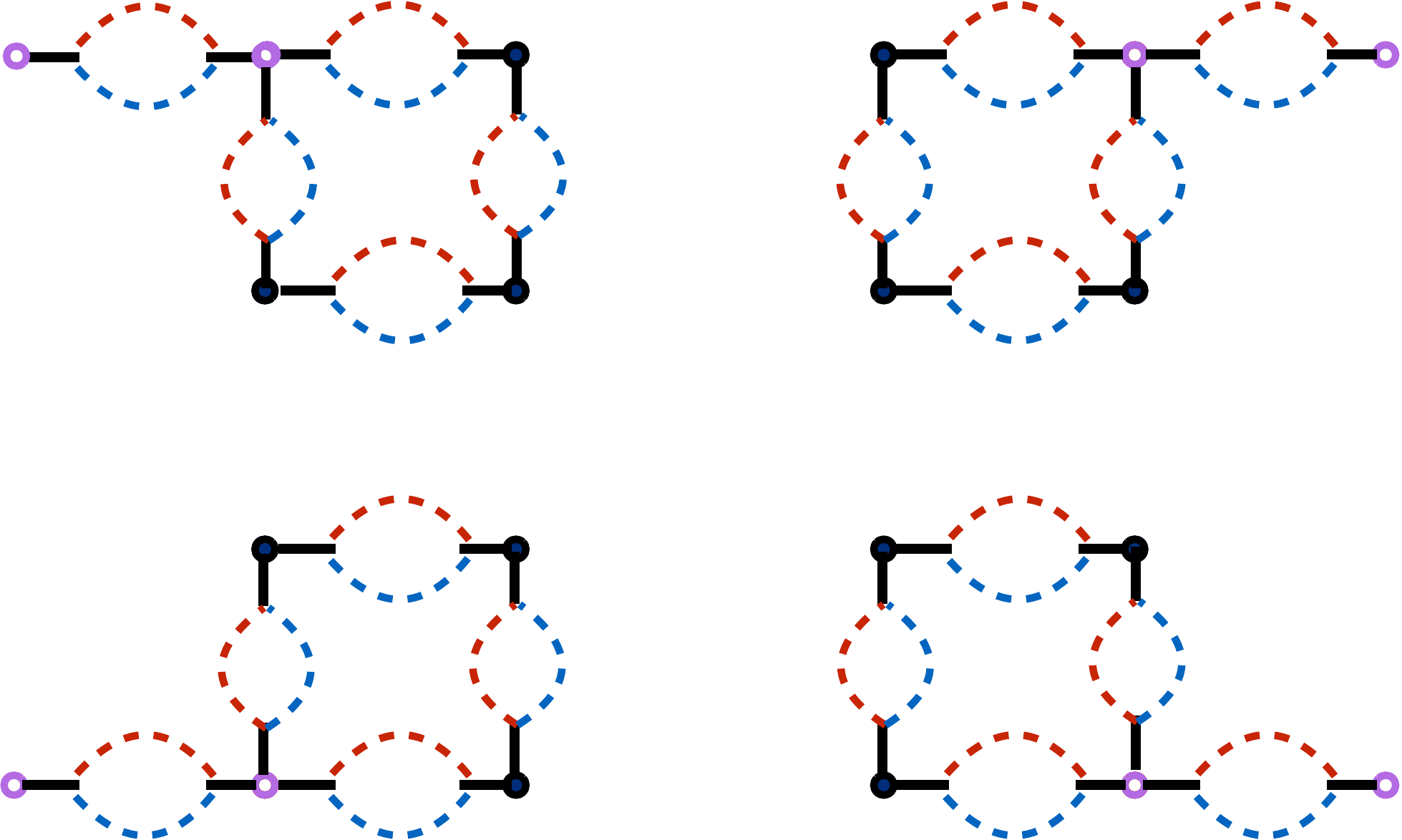}
 %trim option's parameter order: left bottom right top
\caption{(Color online) Fifth-order diagrams for $\rho_{(1,0)}$ (at $h=0$) based on expanding (\ref{sigma_sigma_correlator_1}): These four diagrams involve a three-leg end vertex. Each diagram contributes $(-2)\zeta^5$.}
\label{fig:rho_generic_b1}
\end{figure}
%%%%%%%%%%%%%%%%%%%%%%%%%%%%%%%%%%%%%%%%%

 %%%%%%%%%%%%%%%%%%%%%%%%%%%%%%%%%%%%%%%%%
\begin{figure}[tbp]
\centering
 \includegraphics[trim = 0 0mm 0mm 0, clip, width=1.0\columnwidth]{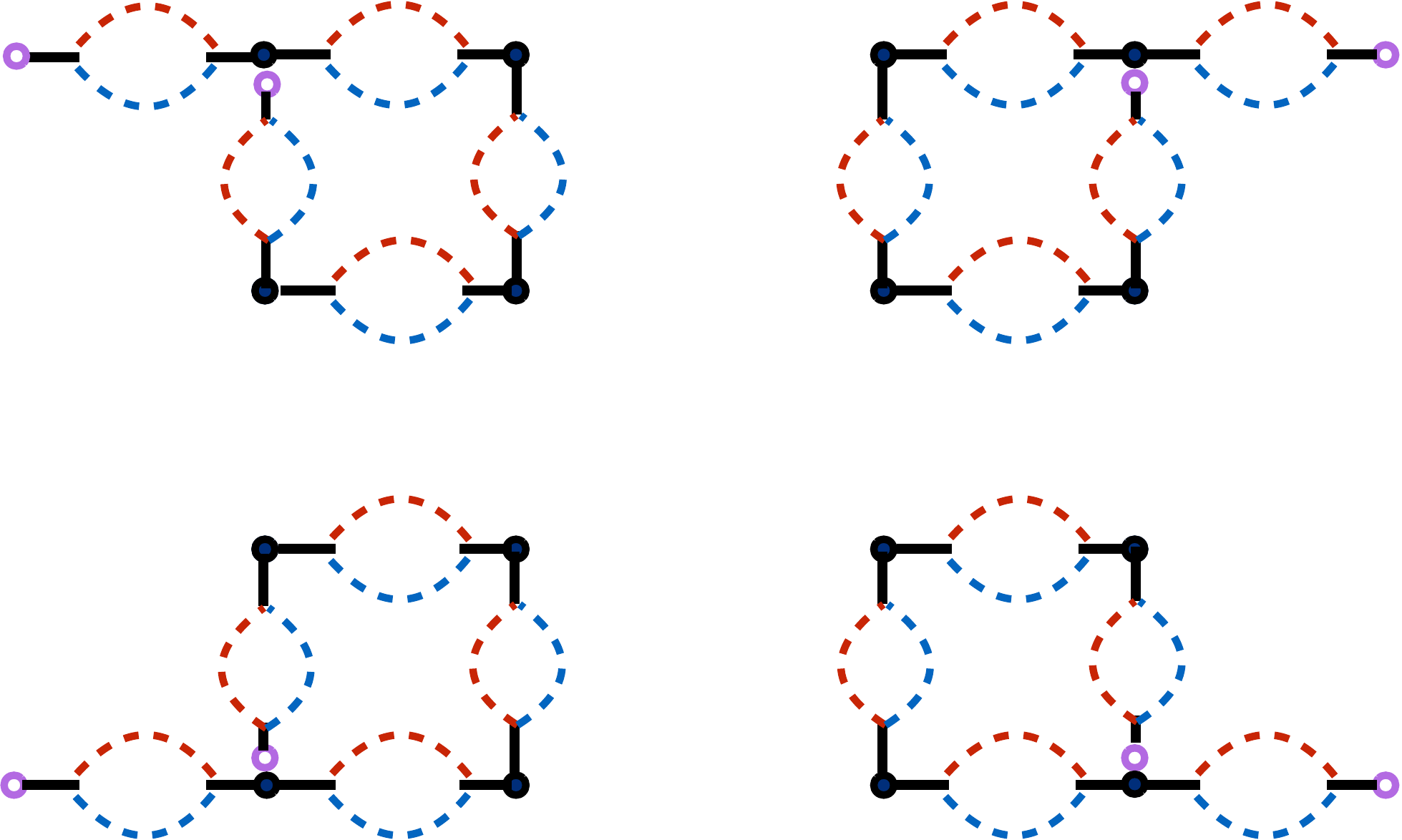}
 %trim option's parameter order: left bottom right top
\caption{(Color online) Fifth-order diagrams for $\rho_{(1,0)}$ (at $h=0$) based on expanding (\ref{sigma_sigma_correlator_1}): These four counterparts of the diagrams shown in 
Fig.~\ref{fig:rho_generic_b1} are obtained by replacing a three-leg end vertex with a one-leg end vertex.
Each diagram contributes $\zeta^5$.}
\label{fig:rho_generic_b2}
\end{figure}
%%%%%%%%%%%%%%%%%%%%%%%%%%%%%%%%%%%%%%%%%

 %%%%%%%%%%%%%%%%%%%%%%%%%%%%%%%%%%%%%%%%%
\begin{figure}[tbp]
\centering
 \includegraphics[trim = 0 0mm 0mm 0, clip, width=1.0\columnwidth]{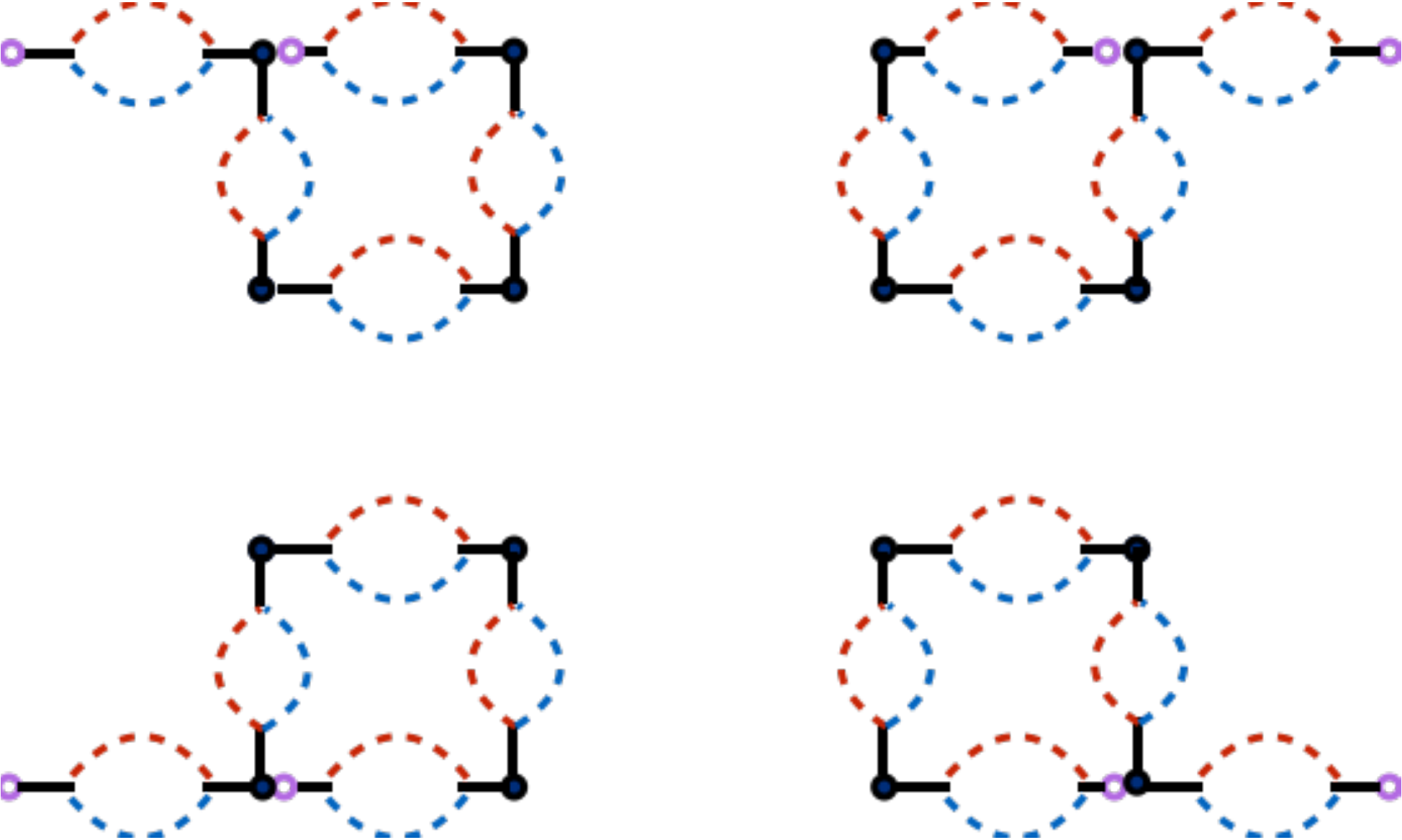}
 %trim option's parameter order: left bottom right top
\caption{(Color online) The four remaining counterparts (cf. Fig.~\ref{fig:rho_generic_b2}) to Fig.~\ref{fig:rho_generic_b1}. }
\label{fig:rho_generic_b}
\end{figure}
%%%%%%%%%%%%%%%%%%%%%%%%%%%%%%%%%%%%%%%%%

 %%%%%%%%%%%%%%%%%%%%%%%%%%%%%%%%%%%%%%%%%
\begin{figure}[tbp]
\centering
 \includegraphics[trim = 0 0mm 0mm 0mm, clip, width=1.0\columnwidth]{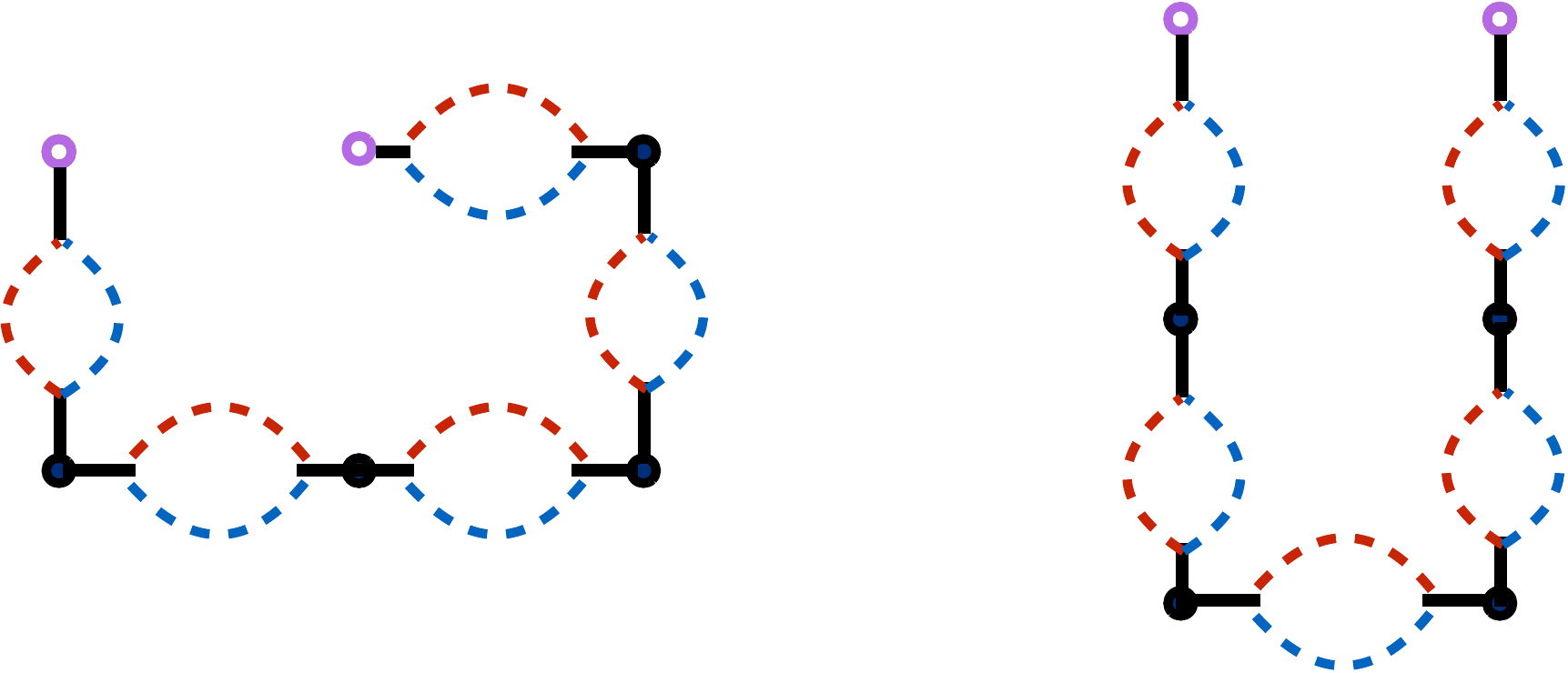}
 %trim option's parameter order: left bottom right top
\caption{(Color online) Additional fifth-order diagrams for $\rho_{(1,0)}$ (at $h=0$) involving two one-leg end vertexes.  Each diagram contributes $\zeta^5$.}
\label{fig:rho_generic_c}
\end{figure}
%%%%%%%%%%%%%%%%%%%%%%%%%%%%%%%%%%%%%%%%%

%%%%%%%%%%%%%%%%%%%%%%%%%%%%%%%%%%%%%%%%%
\begin{figure}[tbp]
\centering
 \includegraphics[trim = 0 0mm 0mm 0mm, clip, width=0.6\columnwidth]{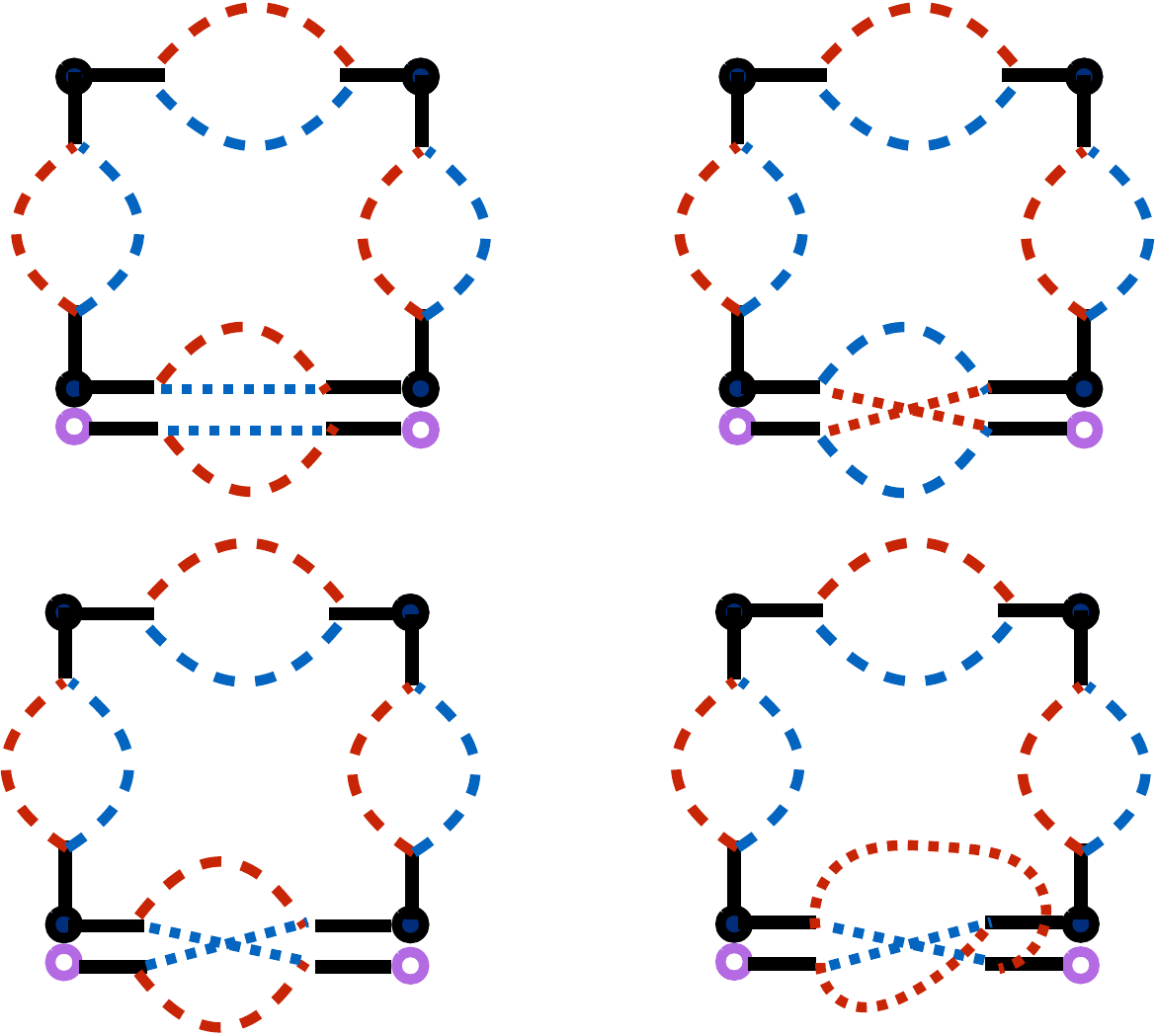}
 %trim option's parameter order: left bottom right top
\caption{(Color online) Fifth-order diagrams for $\rho_{(1,0)}$ (at $h=0$)  containing a link with multiple Grassmann pairs. The net sum of the shown diagrams is $- \zeta^5$, because there are three ways of associating the primed and non-primed propagators along the bottom link,
two of them contribute with the negative sign (upper right and lower left panel) and the third one is contributing with the positive sign (lower right panel). The remaining possibility
(shown in the upper left panel) is not allowed since it produces a disconnected diagram.}
\label{fig:rho_generic_d}
\end{figure}
%%%%%%%%%%%%%%%%%%%%%%%%%%%%%%%%%%%%%%%%%

\subsection{Model and observables}

Consider the 2D Ising model on the square lattice with the Hamiltonian
\begin{equation}
-H/T = \beta \sum_{\langle i,j \rangle} \sigma_i \sigma_j + \sum_i h_i \sigma_i.
\end{equation}
The Ising variables $\sigma = \pm 1$ live on the sites of the 2D square lattice and interact ferromagnetically with their nearest neighbors, as is represented by the first term in the Hamiltonian. We write the dimensionless coupling as $\beta$ in units of the temperature $T$.
Additionally, every spin feels a dimensionless magnetic field $h_i=h$, which can be taken
$h \ge 0$ without loss of generality.
The partition function of the Ising model reads
\begin{equation}
Z = \sum_{ \{ \sigma_i \} }
\prod_{\langle i,j \rangle} e^{\beta \sigma_i \sigma_j} \;  \prod_{i} e^{h_i \sigma_i}.
\end{equation}
The most typical observable of the Ising model is the spin-spin correlation function $\rho_{ij}$,
\begin{equation}
\rho_{\ij} = \langle \sigma_i \sigma_j \rangle = \frac{1}{Z} \left.
\frac{ \partial^2 Z}{\partial h_i \partial h_j} \right|_{h_i = h_j= h} \,.
\label{sigma_sigma_correlator_1}
\end{equation}

\subsection{Grassmannization of the high-temperature expansion}

Using the well-known identities
\begin{eqnarray}
e^{\beta \sigma_i \sigma_j} & = & \cosh \beta \left( 1 + \sigma_i \sigma_j \tanh \beta \right) \nonumber \\
e^{h \sigma_i} & = & \cosh h \left(  1 + \sigma_i \tanh h \right),
\end{eqnarray}
the partition function can be written as $Z = Z_0 Z'$ with
$Z_0 = ( \cosh \beta)^{2N} ( \cosh h)^N$ for a lattice of $N$ sites and $2N$ links. With the notation
$\zeta = \tanh \beta$ and $\eta = \tanh h$ the remaining factor is given by
\begin{equation}
Z' = \sum_{ \{ \sigma_i \} } \, \prod_{\langle i,j \rangle} ( 1 + \sigma_i \sigma_j \zeta )
 \, \prod_{i} ( 1 + \sigma_i \eta).
\end{equation}
Upon summation over spin variables
we are left with a link model, where link variables take only two values, 0 or 1, to
specify whether we are dealing with the first or the second term in the sum
$(1 + \sigma_i \sigma_j \zeta)$. In the partition function, terms with an odd power of $\sigma_i$ on any of the sites yield zero upon spin summation. The remaining terms depend on link
variables in a unique way.
The formalism of the previous section can be straightforwardly applied, and we obtain
\begin{eqnarray}
& f(0)  =  1\,, \;\;  f(1)  =   f_* = \zeta \,, \\
& g(0)  =  g(2) = g(4)=1 \;, \;\; g(1) =  g(3)=\eta \,.  \label{eq:factors_f_g}
\end{eqnarray}
Here we label site factors using the total sum of incident link variables,
$\sum_{b \in \{ b \}_j } \alpha_b$, to avoid unnecessary rank-4 tensor notations.
If we further redefine $Z_0 \to Z_0 2^N f_*^{2N}$, then the Grassmann representation
of the partition function $Z'$ is given by
\begin{eqnarray}
Z' &= &  \int \mathcal{D}[ \bar{\xi}' \xi' \bar{\xi} \xi  ]_{ \{ \alpha_b \} } \prod_{ \{ \alpha_b \} } \exp \left( \frac{1}{\sqrt{\zeta}} \bar{\xi}'_{\alpha_b} \xi_{\alpha_b} + \frac{1}{\sqrt{\zeta}} \bar{\xi}_{\alpha_b} \xi_{\alpha_b} \right) \nonumber \\
{} & {} & \times  \exp \left( \sum_j \lambda_{ \alpha_{ \{b \} } } \prod_{b_j} \breve{\xi}_{\alpha_b} \breve{\xi}^*_{\alpha_b}   \right)  \,.
\end{eqnarray}

\subsection{Vertex coefficients}

We now compute the factors $\lambda$. To this end, we first introduce notations
(for a fixed site $j$ and suppressing the site index for clarity)
\begin{eqnarray}
V_1\!\! &=&\!\! \breve{\xi}_R  \breve{\xi}_R^* +  \breve{\xi}_U  \breve{\xi}_U^* + \breve{\xi}_L  \breve{\xi}_L^* + \breve{\xi}_D  \breve{\xi}_D^* = n_R + n_U + n_L + n_D, \nonumber \\
V_2\!\! &=&\!\! n_R n_U + n_R n_L + n_R n_D + n_U n_L + n_U n_D + n_L n_D,  \nonumber \\
V_3\!\! &=&\!\! n_R n_U n_L + n_R n_U n_D + n_R n_L n_D + n_U n_L n_D,  \nonumber \\
V_4\! \!&=&\! \!n_R n_U n_L n_D ,
\end{eqnarray}
and then Taylor expand
\begin{equation}
\exp \left[ \lambda_1 V_1 + \lambda_2 V_2 + \lambda_3 V_3 + \lambda_4 V_4 \right].
\end{equation}
The only non-zero terms generated by this expansion are $V_1^2 = 2V_2, V_1^3 = 6V_3, V_1^4 = 24 V_4,  V_1 V_2 = 3V_3, V_1 V_3 = 4 V_4$ and $V_2^2 = 6 V_4$. All other powers and multiplications
of operators yield zero. Note that operators from different sites commute and may be
excluded from consideration here. The final result is
\begin{eqnarray}
\lefteqn{\exp \left[ \lambda_1 V_1 + \lambda_2 V_2 + \lambda_3 V_3 + \lambda_4 V_4 \right] = } \nonumber  \\
& & 1 + \lambda_1 V_1 + \lambda_2 V_2 + \lambda_3 V_3 + \lambda_4 V_4  \nonumber \\
& & + \frac{1}{2} \left( \lambda_1^2 2V_2 + \lambda_2^2 6V_4 + 2 \lambda_1 \lambda_2 3 V_3 + 2 \lambda_1 \lambda_3 4V_4 \right) \nonumber \\
& & + \frac{1}{6} \left( \lambda_1^3 6V_3 + 3 \lambda_1^2 \lambda_2 12 V_4 \right) + \frac{1}{24} \lambda_1^4 24 V_4 .
\end{eqnarray}
Term-by-term matching with Eq.~(\ref{eq:factors_f_g}) then leads to
\begin{eqnarray}
g_1  = \eta & = & \lambda_1\, , \\
g_2  = 1    & = & \lambda_2 + \lambda_1^2\, ,\\
g_3  = \eta & = & \lambda_3 + 3 \lambda_1 \lambda_2 + \lambda_1^3\,  ,\\
g_4  = 1    & = & \lambda_4 + 3 \lambda_2^2 + 4 \lambda_1 \lambda_3 + 6 \lambda_2 \lambda_1^4 + \lambda_1^4 \, .
\end{eqnarray}
The solution is immediate
\begin{eqnarray}
\lambda_1 & = & \eta \, , \\
\lambda_2 & = & 1 - \eta^2\,  ,  \\
\lambda_3 & = & -2 \eta + 2 \eta^3\, , \\
\lambda_4 & = &  -2 + 8 \eta^2 - 6 \eta^4 \, .
\end{eqnarray}

In what follows we will discuss the $\eta=0$ case (zero external field)
when the only vertexes with non-zero coupling in the partition function
are $V_2$ and $V_4$,
\begin{equation}
\prod_j \exp(V_2^{(j)} + V_4^{(j)}) = \exp \left\{ \sum_j (V_2^{(j)} - 2 V_4^{(j)}) \right\}.
\end{equation}
The expansion of $Z'$ in powers of $\zeta$ then goes as
\begin{eqnarray}
Z' & = & \prod_b \int \mathcal{D} [...]_b \exp \left( \frac{1}{\sqrt{\zeta}} \bar{\xi}'_b \xi'_b + \frac{1}{\sqrt{\zeta}} \bar{\xi}_b  \xi_b \right) \nonumber \\
{} & {} & \times \exp \left\{ \sum_j (V_2^{(j)} - 2 V_4^{(j)}) \right\}
  \nonumber \\
&  = & \left[ 1 +  4 \zeta^4 + 12 \zeta^6 + \ldots \right]^N.
\label{zetprim}
\end{eqnarray}
and the spin-spin correlation function is given by
\begin{eqnarray}
\rho_{ij} & = & \frac{1}{Z'}\prod_b \int \mathcal{D} [...]_b \exp \left( \frac{1}{\sqrt{\zeta}} \bar{\xi}'_b \xi'_b + \frac{1}{\sqrt{\zeta}} \bar{\xi}_b  \xi_b \right) \nonumber \\
{} & {} & \times (V_1^{(i)} - 2 V_3^{(i)})(V_1^{(j)} - 2 V_3^{(j)}) .
\label{rhoprim}
\end{eqnarray}

\subsection{Feynman rules}

In order to arrive at the Feynman perturbative expansion we need to write the partition function in the form
\begin{equation}
Z' = Z_{\cal l} \left( \sum_{n=0}^{\infty} \sum_{x_1, \ldots, x_n}  \frac{(+1)^n}{n!} \langle V(x_1) \ldots V(x_n) \rangle_0 \right) ,
\end{equation}
where $Z_{\cal l}$ is the partition function of the Gaussian part
(it is the product of local link contributions),
$Z_{\cal l} = \prod_b \int  \mathcal{D} [...] \exp( \bar{\xi}'_b \xi'_b + \frac{1}{\zeta} \bar{\xi}_b \xi_b )   = (1 + \zeta)^{(2N)}$.
%Our 'Heisenberg' operators $V(x_1)$ we conventionally construct by assigning the $H_0$ parts on the $+x$ and $+y$ directions. Note that when we make this step, we leave the high-temperature interpretation that every lattice link can be occupied at most once, since we now have the possibility of multiple vertexes per lattice site.
Feynman rules for the correlation function of the 2D Ising model now follow from the textbook
considerations:
\begin{enumerate}
\item The bare propagators $G^{(0)}=\sqrt{\zeta}$ for primed and non-primes variables are local and reside on the links of the original lattice.
In the correlation function they always occur in pairs of conjugate Grassmann variables
and each pair contributes a factor $\zeta$.
The propagation lines do not have arrows. The bare interaction vertexes (or pre-diagrams, see Fig.~\ref{fig:prediagrams}) are also local and live on the sites of the lattice. There are different types belonging to the  $V_2$ and $V_4$ classes with weight $1$ and $-2$, respectively [see Eq.~(\ref{zetprim})]. On the first (and last) site of the correlator we have a vertex belonging to the class $V_1$ or $V_3$ (see Figs.~\ref{fig:rho_generic_a}-\ref{fig:rho_generic_c}) with weight $1$ and $-2$, respectively [see (\ref{rhoprim})].
\item Draw in order $n$ all topologically distinct connected diagrams with $n$ pairs of bi-grassmann variables living on the links of the lattice. The number of interaction vertexes, excluding the end points, is at most $n-1$.
\item For links with multiple occupancy, a minus sign occurs when swapping 2 Grassmann variables. The minus sign can also be found by counting all closed fermionic loops.
%\item A symmetry factor $s=1/m!$ is assigned when the same vertex is found $m$ times on the same site (see below).
%\item The total weight of the diagram in order $n$ is hence $(-1)^P (-2)^q \zeta^n/s$ with $P$ the signature of the permutation, $q$ the sum of all type-3 and type-4 vertexes, and $s$ the symmetry factor.
\item The total weight of the diagram in order $n$ is hence $(-1)^P (-2)^q \zeta^n$ with $P$ the signature of the exchange permutation and $q$ the sum of all type-3 and type-4 vertexes. %, and $s$ the symmetry factor.
\end{enumerate}
Disconnected diagrams are defined with respect to both the primed and non-primed Grassmann variables simultaneously. Thus, a link can lead to a disconnected diagram only if the primed and non-primed variables simultaneously lead to disconnected pieces (such as the upper left panel in Fig.~\ref{fig:rho_generic_d}). We check the connectivity of a diagram by the breadth-first algorithm.\\

\subsection{Example: the first element of the spin correlation function}

Let us focus on the first element of the correlation function connecting the sites $(0,0)$ and $(1,0)$ (using translational invariance, any 2 neighboring sites $\left< \bm{r}_1, \bm{r}_2 \right>$ can be taken).
To first order, we put a $V_1$ vertex on the origin and target site. There is one way to combine them, thus the total contribution is $\zeta$.
By the symmetry of the lattice, even expansion orders do not contribute. In third order, we can construct a diagram by putting a $V_2$ (RD) vertex on the site $(0,1)$ and a $V_2$ vertex $(LD)$ on the site $(1,1)$. The mirror symmetry of this diagram about the x-axis is also a valid diagram. Hence, the contribution is $2 \zeta^3$. These diagrams contributing in first and third order are shown in Fig.~\ref{fig:rho_generic_a}.

In fifth order, there are 4 diagrams with a $V_3$ vertex on one of the endpoints, yielding a contribution $-8\zeta^5$ . There are 14 diagrams consisting of only $V_1$ and $V_2$ vertexes and single pair-lines, yielding a contribution $14 \zeta^5$. The contributions to fifth order are shown in 
Figs.~\ref{fig:rho_generic_b1},~\ref{fig:rho_generic_b2},~\ref{fig:rho_generic_c}, and~\ref{fig:rho_generic_d}.
There are however additional diagrams with 2 pairs of Grassmann variables living on the same link, as is shown in Fig.~\ref{fig:rho_generic_d} (there are equivalent diagrams obtained by mirror symmetry around the x-axis which are not shown).
%We consider the diagram shown in the lower right panel as the parent diagram.
They all have on the origin a $V_1$ and a $V_2$ ($RU$) vertex, and on the target site $(1,0)$ a $V_1$ and a $V_2$ ($UL$) vertex. On the site $(1,1)$ there is a $V_2$ ($LD$) and on site $(0,1)$ a $V_2$ ($RD$) vertex. Let us look more carefully at the link between the origin and target site:
\begin{equation}
\frac{1}{\zeta^2} \int \mathcal{D} \left[ \bar{\xi}' \bar{\xi} \xi \xi'  \right] \bar{\xi}' \bar{\xi} \xi \xi' \bar{\xi}' \bar{\xi}  \xi \xi' .
\label{example_term}
\end{equation}
The origin is associated with $\bar{\xi}' \bar{\xi}$ and the target with $  \xi \xi'$ by our convention.
Applying Wick's theorem, there are 4 possible ways to pair the Grassmann variables:
\begin{enumerate}
\item The pairing combination
$
\overbracket{\bar{\xi}'\underbracket{\bar{\xi} \xi} \xi'} \overbracket{\bar{\xi}'
\underbracket{\bar{\xi}  \xi} \xi'}
$
comes with the sign +1 and leads to a connected diagram (this is the lower right panel in Fig.~\ref{fig:rho_generic_d}).
\item The pairing combination
$
\mathrlap{\overbracket{\phantom{\bar{\xi}' \bar{\xi} \xi \xi'}}} \bar{\xi}'\underbracket{\bar{\xi} \underbracket{\xi \xi' \mathrlap{\overbracket{\phantom{ \bar{\xi}' \bar{\xi}  \xi \xi'}}}
\bar{\xi}' \bar{\xi}}  \xi} \xi'
$
comes with the sign -1 and leads to a connected diagram (this is the upper right panel in Fig.~\ref{fig:rho_generic_d})
\item The pairing combination
$
\overbracket{\bar{\xi}'\underbracket{\bar{\xi} \xi} \underbracket {\xi'\bar{\xi}'}
\underbracket{\bar{\xi}  \xi} \xi'}
$
comes with the sign -1 and leads to a connected diagram (this is the lower left panel in Fig.~\ref{fig:rho_generic_d})
\item The pairing combination
$
\overbracket{\bar{\xi}'\overbracket{\bar{\xi}\underbracket{\xi \underbracket{\xi'\bar{\xi}'}
\bar{\xi}} \xi} \xi'}
$
leads to a disconnected diagram and does not contribute to the correlation function (this is the upper left panel in Fig.~\ref{fig:rho_generic_d}).

\end{enumerate}
The net contribution of these 4 distinct diagrams is hence $-1$ (also the diagrams obtained by mirror symmetry around the x-axis yield $-1$, so the total contribution to fifth order is $ (-8 + 14 - 2) \zeta^5 = 4 \zeta^5$.

It is instructive to notice that the sum of all diagrams in which multiple Grassmann pairs live on the same link always produces zero in case all diagrams are connected, in line with the nilpotency of Grassmann variables. Wick's theorem splits however these contributions in connected and disconnected diagrams, where the disconnected diagrams cancel against the denominator of the Feynman expansion.  
It is this non-trivial regrouping imposed by Wick's theorem that can yield non-zero 
contributions from terms like (\ref{example_term}); and, in particular, from arbitrarily high powers of one and the same interaction vertex.

%It is this non-trivial regrouping imposed by Wick's theorem that can yield non-zero contributions in terms like (\ref{example_term}) for the correlation function.

\section{Implementation}
\label{sec:V}

We explored two ways of evaluating the (bare) series for the spin-correlator: a stochastic Monte Carlo approach and a deterministic full evaluation of all diagrams.

\subsection{Monte Carlo sampling }

%We sample over all Feynman diagrams by considering a tail and head as endpoints of the correlation function.
%We refer to the endpoints of the correlation function as the head and the tail. By moving around the head, allowing to change the vertex types (which are generated when moving the head), as well as changing the edge connections between the vertices, we will sample over all Feynman diagrams according to the following algorithm:
In order to perform a Monte Carlo sampling over all Feynman diagrams, we introduce a head and a tail that represent the endpoints of the correlation function. By moving them around the lattice and changing the diagrammatic elements in between the head and tail, we are able to reach an ergodic sampling. The algorithm can be formulated as follows:
The tail remains stationary at the origin whereas the head can move around the lattice.  When the head and tail are on the same site and the expansion order is 0, the value of the correlation function is 1 which can be used for normalization of the Monte Carlo process. A Monte Carlo measurement contributes $+1$ or $-1$ depending on the sign of the diagram weight. The simplest Monte Carlo procedure samples according to the absolute weights of the diagrams and consists of the following pairs of reciprocal updates:
\begin{enumerate}
\item MOVE--RETRACT. We choose one of the 4 directions randomly, and attempt to place the head on the site adjacent to the current head site according to this direction. In case this direction does not correspond to backtracking, the current $V_1$ type of the tail turns into a $V_2$, otherwise the head goes back and changes the previous $V_2$ into a $V_1$ type (unless the diagram order is 0 or 1, when only $V_1$ types are possible).
When moving forward, the way of pairing primed and non-primed variables is always unique which in turns implies that we can only retract when the head is connected via a ``straight pair connection" to the previous vertex (both primed and non-primed Grassmann variables of the head are connected to the same vertex on the previous site).  We only allow the MOVE--RETRACT updates if the end vertex types are $V_1$.
\item SWAP VERTEX. Swaps between the vertexes $V_1 + V_2 \leftrightarrow V_3$ (for head and/or tail) and $V_2 + V_2 \leftrightarrow V_4$ (anywhere in the diagram). This update is its own reciprocal.
\item RELINK. On a given link, relink primed and non-primed Grassmann variables. This can change the sign of the weight only. This update is its own reciprocal.
\end{enumerate}
The second and third type of updates may lead to disconnected diagrams. In such cases, the configuration is unphysical. We opt to allow such configurations, but a Monte Carlo measurement is forbidden and type-1 updates remain impossible until the diagram is connected again.
For small values of $\zeta$ the sign problem is nearly absent, but only low expansion orders can be reached. For higher values of $\zeta$ (close to and above the critical one) an increasing number of orders contributes significantly, consequently more time is spent in higher orders and the sign problem significantly worsens.

\subsection{Deterministic full evaluation}

For the case of the 2D Ising model, a Monte Carlo approach offers no advantages over a full series expansion approach. With this we mean the explicit listing and evaluation of all possible diagrams as opposed to the stochastic sampling over all topologies. This is because all diagrams in a given expansion order contribute a number of order unity (times the same power of $\zeta$), often with alternating sign, leading to huge cancellations. Only the exact cancellation has physical information, and this requires that every diagram is evaluated multiple times before the correct convergence can be seen. A Monte Carlo approach makes much more sense if the dominant contributions to the total weight are coming from a narrow parameter region, which is usually the case if there are additional integrals over internal momenta.

We therefore wrote a code that evaluates all diagrams for the correlation function up to a maximum order. The construction is based on the fact that there is an easy way to construct all the ``easy" diagrams (the ones that formally look like originating from a high-temperature series expansion). These can serve as parent diagrams, from which further offspring diagrams can be constructed which have one or multiple $V_3$ and $V_4$ vertexes as well as possible fermionic exchanges. All diagrams in order $n$ can be found as follows:
\begin{enumerate}
\item Write down all possible words of the form $X_1 X_2 \ldots X_n$ with the alphabet $X_j \in \{0,3 \}$ corresponding to the 4 directions on the square lattice. Make sure that subsequent directions are not backtracking. For example, if $X_4$ is in the positive $+\hat{x}$ direction, then $X_5$ cannot be in the negative $-\hat{x}$ direction. From this word we also know all sites and links that are visited, as well as all type-1 and type-2 vertexes that are used to make this diagram.
\item Such a parent diagram is added to a list of different topologies only if it has a unique topology. To store the topological information of a bare vertex, we need to store a pair consisting of a site index and a vertex type. The diagram is then stored as an ordered map where the ``key" values are given first by the lattice site index and second by the vertex type (in binary format). The ordered map may have multiple entries with the same key if multiple vertexes reside on the same site and if they are of the same type ({\it e.g.}, two $RL$ vertexes on the same site).
\item We iterate over this configuration list and check if the tail and head sites can be merged into a type-3 vertex by combining them with type-2 vertexes that reside on the same lattice site. If so, and if the resulting topology is unique, the diagram is added to the list. This step is performed in three parts: first for the head and tail together (in order to find all diagrams with 2 $V_3$ ends), then for the head alone, and finally for the tail alone.
\item We iterate again over the full configuration list and check if 2 type-2 vertexes that live on the same site can be merged into a type-4 vertex. This last step has to be repeated until no further merges are possible (since it may happen that a diagram has multiple type-4 vertexes or even multiple type-4 vertexes on the same site). Diagrams thus created are also added the configuration list if their topology is unique. After completion of this step, all possible topologies have been generated.
\item We compute the product of all the vertex weights, according  to the Feynman rules.
\item From this list of parent diagrams we need to generate all offspring diagrams which feature all possible fermionic permutations for multiply occupied links. This first requires that we know how the vertexes are connected in the parent diagram, which is stored in the configuration list. The parent diagram always has permutation sign +1 (because the connections of the primed and non-primed Grassmann variables are always the same). Next we generate all possible permutations by relinking the primed and/or the non-primed Grassmann variables using Heap's algorithm. If a link has occupation number $m$, then there are $(m!)^2$ combinations to be generated (and there may be more than one multiply occupied link). The permutation signature is also stored.
\item We check the connectivity of the diagram using the breadth-first algorithm. Disconnected diagrams contribute 0.
\item Finally, we compute the isomorphism factor: if $m$ identical vertexes on the same site are found, a factor $1/m!$ must be taken into account. This is a consequence of how we construct the diagrams: topology checks were only performed on the parent diagrams (and based on vertexes only), not on offsprings obtained by fermionic exchange. (It would be prohibitively expensive to add the offspring diagrams to the list of all possible diagrams.) Hence, just as we generate illegal disconnected diagrams, we also have a double counting problem when identical vertexes occur in the list.
\end{enumerate}

In order 14, there were about 140,000 parent diagrams contributing to the first entry on the diagonal of the correlator. The hugest number of permutations was $(4!)^4(3!)^4 \approx 10^8$. Since the sum of these permutations has a net contribution of order 1, Monte Carlo has roughly a sign problem of the order of $10^{-8}$ for these diagrams. The first time a nontrivial isomorphism factor is seen is in order 6 for the first element on the diagonal of the spin correlator: There are diagrams in which two links are doubly occupied, and those links are connected by an identical $V_2$ vertex, hence the isomorphism factor $1/2$. More efficient ways of evaluating and storing the diagrams can probably be devised and implemented, but the above scheme is sufficient to check the validity of the technique and study the transition. \\

%For the self-energy we proceeded in a similar fashion. Now we remove the type-1 and type-3 vertexes from the possibilities, and only consider type-2 or type-4 vertexes. A valid diagram begins and ends with an interaction vertex, with 2 remaining unpaired (amputated) legs in total. The 1p-reducibility was checked by cutting all possible $G_0$ lines of the diagram({\it i.e.,} the outgoing legs of a vertex are connected by the same pair of Grassmanns to the incoming legs of the next vertex), and applying the breadth-first algorithm to check the connectivity.

\section{Results}
\label{sec:VI}

\subsection{Spin-spin correlation function}

Our results for the spin-spin correlation function are shown in Table.~\ref{table:rho}. The correlation function is known recursively from Refs.~\onlinecite{Perk1980, Guttmann_rho1, Guttmann_rho2}. It is also known as a Painlev{\'e}-VI nonlinear differential equation\cite{McCoy76} but this is not so well suited to obtain the series coefficients. Along the principal axes and the diagonal it can also be expressed as a Toeplitz determinant. 
The first element along and the axis and the diagonal can be recast in terms of complete elliptic integrals (see pp. 200-201 in Ref.~\onlinecite{mccoy1973}), which are convenient for series expansions,
\begin{eqnarray}
\rho_{(1,0)}  & = & \coth(2 \beta) \left[ \frac{1}{2} + \frac{\cosh^2 2\beta }{\pi} (2 \tanh^2 2\beta - 1) K(k_>)  \right]  \nonumber \\
{} & \to &  \zeta + 2 \zeta^3 + 4 \zeta^5 + 12 \zeta^7 + 42 \zeta^9 + \ldots \\
\rho_{(1,1)} & = & \frac{2}{\pi k_>} \left[ K(k_>) + (k_>^2 - 1) K(k_>) \right] \\
{} & \to &  2\zeta^2 + 4\zeta^4 + 10 \zeta^6 + 32 \zeta^8 + 118 \zeta^{10} + \ldots \nonumber
\end{eqnarray}
with $k_> = \sinh^2(2\beta)$, $K(.)$ and $E(.)$ the complete elliptic K and E functions, respectively.
The above-cited recursion relations could be initialized with these expansions and shown to yield the same results as the top 2 rows in Table.~\ref{table:rho}.

\begin{table}
\begin{tabular}{|c|c|c|c|c|c|c|c|c|c|c|c|}
\hline
site/order & $\zeta$ & $\zeta^2$ & $\zeta^3$ & $\zeta^4$ & $\zeta^5$ & $\zeta^6$ & $\zeta^7$ & $\zeta^8$ & $\zeta^9$ & $\zeta^{10}$ & $\zeta^{11}$  \\ \hline
(1,0) & 1 & 0 & 2 & 0 & 4 & 0 & 12 & 0 & 42 & 0 & 164 \\ \hline
(1,1) & 0 & 2 & 0 & 4 & 0 & 10 & 0 & 32 & 0 & 118 & 0 \\ \hline
(2,0) & 0 & 1 & 0 & 6 & 0 & 16 & 0 & 46 & 0 & 158 & 0 \\ \hline
(2,1) & 0 & 0 & 3 & 0 & 11& 0 & 31 & 0   & 97  & 0   & 351  \\ \hline
(2,2) & 0 & 0 & 0 & 0 & 6 & 0 & 24 & 76 & 0 & 248 & 0 \\ \hline
(3,0) & 0 & 0 & 1 & 0 & 12 & 0 & 48 & 0 &152 & 0 &506\\ \hline
(3,1) & 0 & 0 & 0 & 4 & 0 & 26 & 0 & 92  & 0  &298  & 0\\ \hline
(3,2) & 0 & 0 & 0 & 0 & 10 & 0 & 55 & 0 &  201 & 0 & 684\\ \hline
(3,3) & 0 & 0 & 0 & 0 & 0 & 20 & 0 & 120 & 0 & 480 & 0 \\ \hline
(4,0) & 0 & 0 & 0 & 1 & 0 & 20 & 0 & 118 & 0 &7 452 & 0\\ \hline
(4,1) & 0 & 0 & 0 & 0 & 5 & 0 & 52 & 0  & 244 & 0  & 885\\ \hline
(4,2) & 0 & 0 & 0 & 0 & 0 & 15 & 0 & 118  &0  & 521 & 0 \\ \hline
(4,3) & 0 & 0 & 0 & 0 & 0 & 0 & 25 & 0  &259  & 0  & 1176 \\ \hline
(4,4) & 0 & 0 & 0 & 0 & 0 & 0 & 0 & 70 & 0 & 560 & 0 \\ \hline
(5,0) & 0 & 0 & 0 & 0 & 1 & 0 & 30 & 0 & 250 & 0 & 1200 \\ \hline
(5,1) & 0 & 0 & 0 & 0 & 0 & 6 & 0 & 92 & 0 & 574 & 0 \\ \hline
(5,2) & 0 & 0 & 0 & 0 & 0 & 0 & 21 & 0 & 231 & 0  & 1266 \\ \hline
(5,3) & 0 & 0 & 0 & 0 & 0 & 0 & 0 & 56 & 0 & 532 & 0 \\ \hline
(5,4) & 0 & 0 & 0 & 0 & 0 & 0 & 0 & 0 & 126 & 0 & 1176 \\ \hline
(5,5) & 0 & 0 & 0 & 0 & 0 & 0 & 0 & 0 & 0 & 252 & 0 \\ \hline
(6,0) & 0 & 0 & 0 & 0 & 0 & 1 & 0 & 42 & 0 & 474 & 0 \\ \hline
(6,1) & 0 & 0 & 0 & 0 & 0 & 0 & 7 & 0 & 149 & 0 & 1215\\ \hline
(6,2) & 0 & 0 & 0 & 0 & 0 & 0 & 0 & 28 & 0 & 416 & 0 \\ \hline
(6,3) & 0 & 0 & 0 & 0 & 0 & 0 & 0 & 0 & 84 & 0 & 1026 \\ \hline
(6,4) & 0 & 0 & 0 & 0 & 0 & 0 & 0 & 0 & 0 & 210 & 0 \\ \hline
(6,5) & 0 & 0 & 0 & 0 & 0 & 0 & 0 & 0 & 0 & 0  & 462 \\ \hline
(7,0) & 0 & 0 & 0 & 0 & 0 & 0 & 1 & 0 & 56 & 0 & 826\\ \hline
(7,1) & 0 & 0 & 0 & 0 & 0 & 0 & 0 & 8 & 0 & 226 & 0 \\ \hline
(7,2) & 0 & 0 & 0 & 0 & 0 & 0 & 0 & 0 & 36 & 0 & 699 \\ \hline
(7,3) & 0 & 0 & 0 & 0 & 0 & 0 & 0 & 0 & 0 & 120 & 0 \\ \hline
(7,4) & 0 & 0 & 0 & 0 & 0 & 0 & 0 & 0 & 0 & 0 & 330 \\ \hline
(8,0) & 0 & 0 & 0 & 0 & 0 & 0 & 0 & 1 & 0 & 72 & 0 \\ \hline
(8,1) & 0 & 0 & 0 & 0 & 0 & 0 & 0 & 0 & 9 & 0 & 326 \\ \hline
(8,2) & 0 & 0 & 0 & 0 & 0 & 0 & 0 & 0 & 0 & 45 & 0 \\ \hline
(8,3) & 0 & 0 & 0 & 0 & 0 & 0 & 0 & 0 & 0 & 0 & 165 \\ \hline
(9,0) & 0 & 0 & 0 & 0 & 0 & 0 & 0 & 0 & 1 & 0 & 90 \\ \hline
(9,1) & 0 & 0 & 0 & 0 & 0 & 0 & 0 & 0 & 0 & 10 & 0 \\ \hline
(9,2) & 0 & 0 & 0 & 0 & 0 & 0 & 0 & 0 & 0 & 0 & 55 \\ \hline
(10,0)& 0 & 0 & 0 & 0 & 0 & 0 & 0 & 0 & 0 & 1 & 0 \\ \hline
(10,1)& 0 & 0 & 0 & 0 & 0 & 0 & 0 & 0 & 0 & 0 & 11 \\ \hline
(11,0)& 0 & 0 & 0 & 0 & 0 & 0 & 0 & 0 & 0 & 0 & 1 \\ \hline
\hline
\end{tabular}
\caption{Expansion coefficients for the correlation function up to order 11.}
\label{table:rho}
\end{table}

\subsection{Magnetic susceptibility}

%%%%%%%%%%%%%%%%%%%%%%%%%%%%%%%%%%%%%%%%%
\begin{figure}[tbp]
\centering
 \includegraphics[width=1.0\columnwidth]{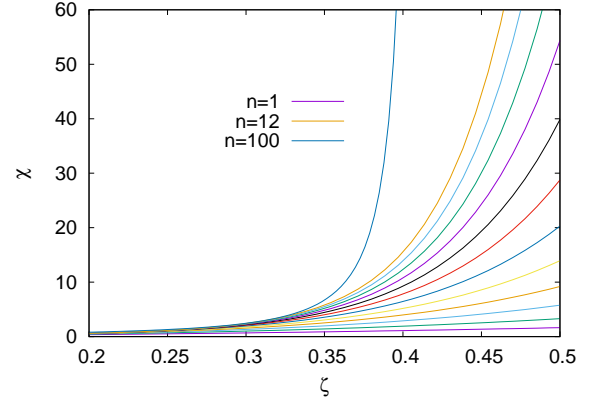}
 %trim option's parameter order: left bottom right top
\caption{(Color online) The magnetic susceptibility versus $\zeta$ for different expansion orders from 12 to 1 (top to bottom), compared 
to the order 100 result---the converged answer over this plotting range---obtained from Ref.~\onlinecite{Guttmann}, 
which shows a divergence in good agreement with the critical exponent $\gamma = 7/4$ starting from $\beta \ge 0.38$}
\label{fig:chi}
\end{figure}
\begin{figure}[tbp]
\centering
 \includegraphics[width=1.0\columnwidth]{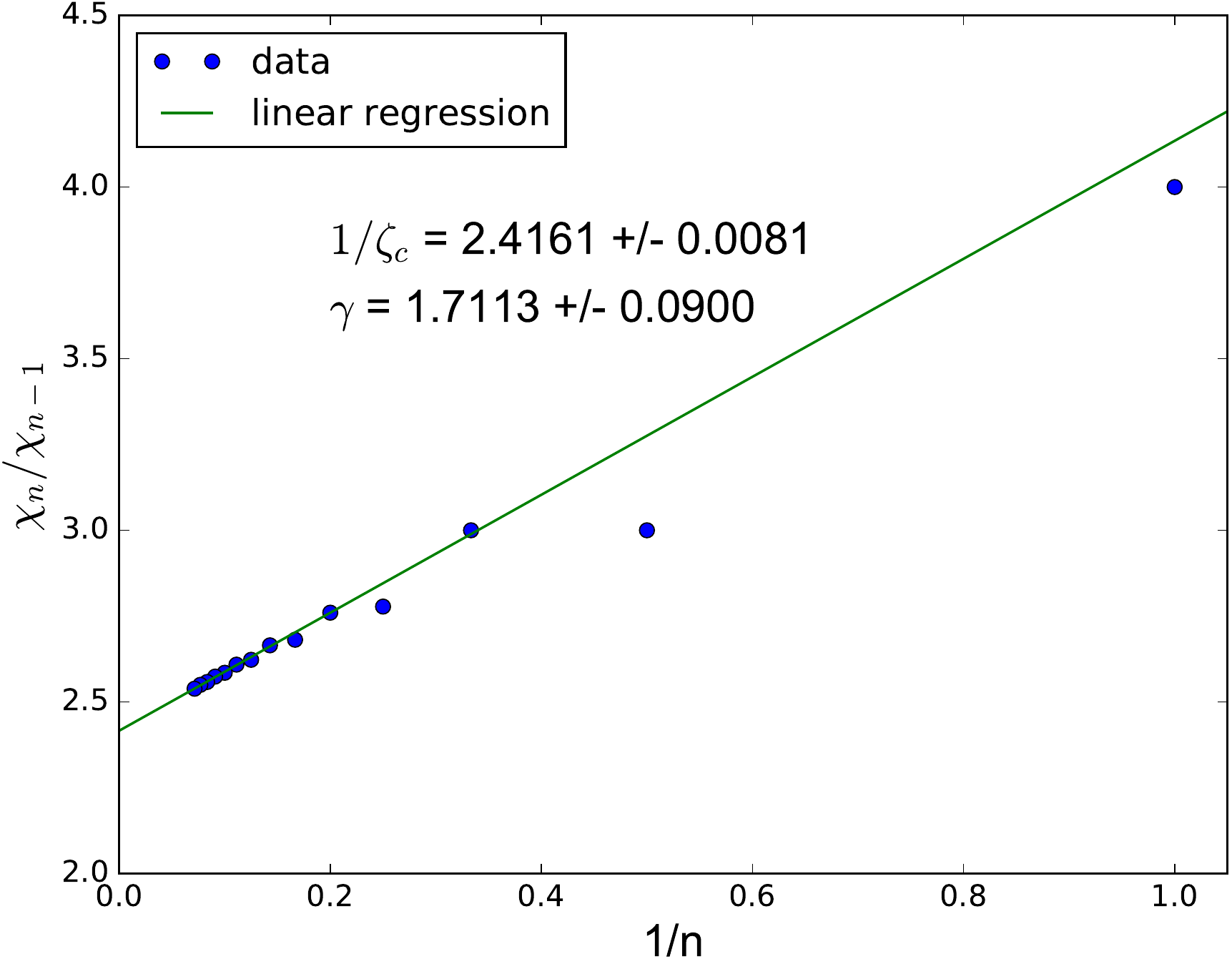}
 %trim option's parameter order: left bottom right top
\caption{(Color online) Ratio of consecutive coefficients $\chi[n-1]$ and $\chi[n]$ in the expansion of the susceptibility as a function of the inverse of the expansion order $1/n$. Linear regression according to Eq.~(\ref{eq:lin_reg}) allows to determine the critical temperature with an accuracy of $0.5\%$ and the critical exponent $\gamma$ with an accuracy of $5\%$. The fitting regime included orders 9 through 14.
 }
\label{fig:chi_ratio}
\end{figure}
%%%%%%%%%%%%%%%%%%%%%%%%%%%%%%%%%%%%%%%%%

The spin susceptibility is related to the zero momentum value of the Green function by $\beta^{-1} \chi = 1 +\rho (\bm{p}=0)$.
We can hence sum over the entire lattice to obtain
%
%
%
%%%%%%%%%%%%%%%%%%%%%%%%%%%%%%%%%%%%%%%%%%%% MISHA START
\begin{eqnarray}
\beta^{-1} \chi & = &  1 + 4 \zeta + 12 \zeta^2 + 36\zeta^3 + 100 \zeta^4 + 276\zeta^5 \nonumber \\
{} & {} & +740 \zeta^6 +1972 \zeta^7  +  5172 \zeta^8 + 13492 \zeta^9 \nonumber \\
{} & {} & + 34876 \zeta^{10} + 89764 \zeta^{11} + 229628 \zeta^{12} \nonumber \\
{} & {} & + 585508 \zeta^{13} + 1486308 \zeta^{14} +\ldots
\end{eqnarray}
To this order the series expansion agrees with the ones from Ref.~\onlinecite{Sykes1972} and Ref.~\onlinecite{Guttmann}.   For a library of high-temperature series expansions, 
see Ref.~\onlinecite{Butera2002}. Currently, the series is known (at least) up to order 2000 and still topic of active research.\cite{Guttmann, Guttmann_rho2}
The series is convergent for any finite expansion order, {\it i.e.}, in the thermodynamic limit the infinite series will diverge first at the phase transition point.
It is hence possible to study the critical behavior of the susceptibility, which is governed by the critical exponent $\gamma = 7/4$.
We plot in Fig.~\ref{fig:chi} the susceptibility versus $\beta$ for different expansion orders, and also plot the asymptotic behavior for comparison.

The critical temperature and the exponent $\gamma$ can be found from a study of the convergence radius of the series. Since
\begin{eqnarray}
\beta^{-1} \chi & = & \sum_n \chi_n \zeta^n \propto (1 -  \zeta/ \zeta_c)^{-\gamma} \\
{} & = & 1 + \sum_{n=1}^{\infty} \frac{ \gamma (\gamma - 1) \cdots (\gamma + n - 1)}{ n!} \left( \frac{\zeta}{\zeta_c} \right)^n \nonumber
\label{eq:lin_reg}
\end{eqnarray}
the ratio of coefficients asympotically behaves as
\begin{equation}
\frac{\chi_n}{\chi_{n-1}} = \frac{1}{\zeta_c} + \frac{\gamma - 1}{\zeta_c}\frac{1}{n}.
\end{equation}
In Fig.~\ref{fig:chi_ratio} we extract the critical point $\zeta_c$ from the intercept and the critical exponent $\gamma$ from a linear fit through the ratio of the coefficients. The critical point could be determined with an accuracy of $0.5\%$, whereas the error on $\gamma$ is of the order of $5\%$. However, according to more advanced extrapolation techniques discussed in 
Ref.~\onlinecite{ZinnJustin1979}, $\gamma$ can be determined independently from $\zeta_c$ as $\gamma \approx 1.751949$ on the square lattice when the series is known up to 14th order, {\it i.e.}, an accuracy of $0.5\%$.

%%%%%%%%%%%%%%%%%%%%%%%%%%%%%%%%%%% KOLYA BELOW %%%%%%%%%%%%%%%%%%%%%%%%%%%%%%%%%%%%%%%%%%

\section{The $G^2W$ skeleton scheme}
\label{sec:VII}

The expansion of susceptibility in terms of $\zeta$ is, of course, identical to the one found by the high-temperature series expansion method. To make the distinction between the high-temperature series formalism and Grassmannization approach clear, we discuss the skeleton formulation of the interacting fermionic field-theory based on dressed (or ``bold") one-body propagators ($G$) and bold interaction lines ($W$). This leads to the so-called $G^2W$ skeleton scheme (see for instance Refs.~\onlinecite{Heidin,Molinari2006} for the terminology): all lines in all diagrams
are assumed to be fully renormalized propagators and effective potentials, but vertex functions
remain bare. In Sec. \ref{sec:VIII} we show that the $G^2W$-expansion scheme offers a very simple
way to solve the 1D Ising model exactly.

\subsection{Objects and notation}

\begin{figure}[tbp]
\centering
 \includegraphics[width=0.3\columnwidth, angle=-90]{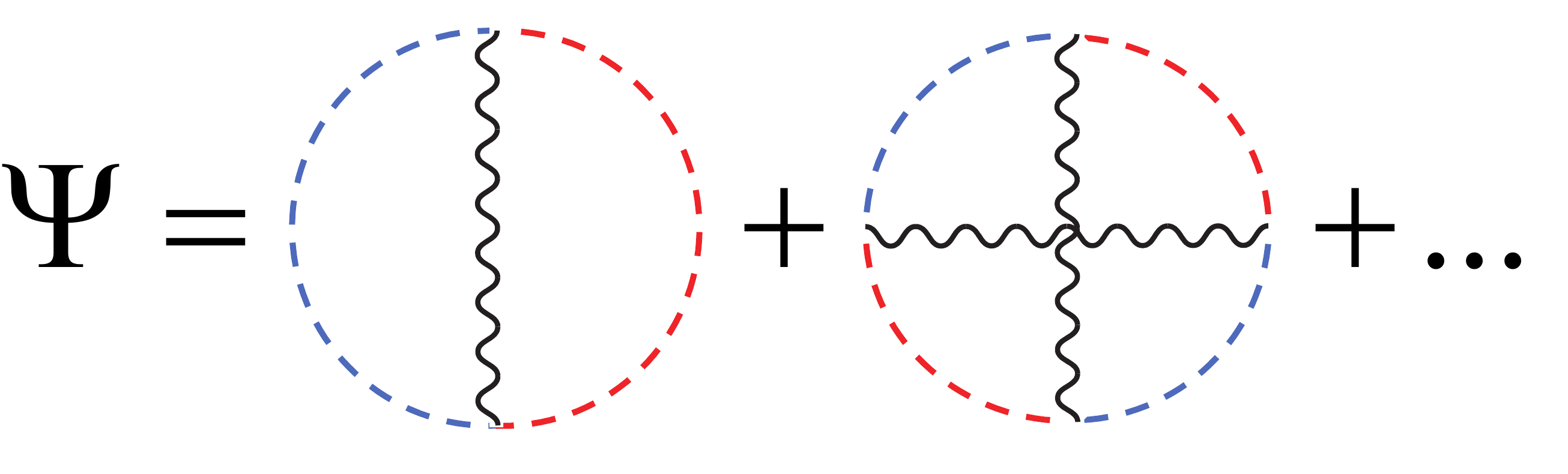}
\caption{Two low-order contributions to the generalized Luttinger-Ward functional $\Psi$.
Dashed lines denote bold Green functions for primed and non-primed Grassmann variables, and
wavy solid lines are effective potential lines.
}
\label{fig:psi_functional}
\end{figure}

The key objects in the standard skeleton scheme are the selfenergy ($\Sigma$) and the polarization function ($\Pi$). They are related to the Green function ($G$) and the
effective potential ($W$) by their respective Dyson equations.
The diagrams for $\Pi$ and $\Sigma$ are obtained by removing one $W$- or $G$-line, respectively,
from connected graphs for the generalized Luttinger-Ward functional $\Psi$, shown to second order in Fig.~\ref{fig:psi_functional}. In this setup, the expansion order is defined by
the number of $W$-lines (obviously, the discussion of Sec. \ref{subsec:D} does not apply
to the self-consistent skeleton sequence). All objects of interest are tensors; they have a coordinate (or momentum) dependence, as well as the legs orientation dependence for the incoming
and outgoing parts. This conventional scheme has to be supplemented with $\Psi$-graphs
involving $V_4$ vertexes to account for all contributions. We start with neglecting $V_4$ vertexes,
and discuss their role later.

In more detail, the formalism of the $G^2W$ expansion in the absence of $V_4$ vertexes
is as follows:
\begin{enumerate}
\item There are six bare two-body interaction vertexes $V_2$ $( RU, RL, RD, LU, UD, LD)$, see the second line in Fig.~\ref{fig:prediagrams}. They reside on the sites of the original square lattice and all have weight 1. Symbolically, we encode the tensor structure
    of $V_2$ using a convenient short hand notation
    $V_2=\sum_{\alpha , \gamma =1}^{4} V(\alpha, \gamma )n_{\alpha} n_{\gamma}$, where
\begin{equation}
V(\alpha, \gamma) = \begin{bmatrix} 0 & 1 &1 & 1 \\ 1 & 0 & 1 & 1 \\ 1 & 1 & 0 & 1 \\ 1 & 1 & 1 & 0 \end{bmatrix} .
\end{equation}
The row index represents the first leg enumerated according to the convention
$(R,U,L,D) \to (0,1,2,3)$, and the column index represents the second leg.
By doing so, we artificially double the number of vertexes from 6 to 12.
For example, the element $(0,2)$ corresponds to $n_Ln_R$ whereas $(2,0)$
corresponds to $n_Rn_L$, which is exactly the same term.

\item The selfenergies $\Sigma$ for the primed and non-primed Grassmann variables take the same value. Thus, we have to compute only one of them and we can suppress the index that distinguishes between the two Grassmann fields. The selfenergy defines the Green function
    through the Dyson equation
\begin{equation}
G(\alpha, \gamma ) = G^{(0)} (\alpha, \gamma ) + \sum_{\mu, \nu}
G^{(0)}(\alpha, \mu ) \Sigma(\mu, \nu ) G(\nu, \gamma ) \,. \label{eq:Dyson}
\end{equation}
For a link going from site $i$ to site $j$, the first index $\alpha$ refers to site $i$ (in the above-defined sense), and the second index $\gamma$ refers to site $j$.
% The convention for the matrix of the legs is hence the transpose of the matrix for $W$ or $V$,
% which was defined with respect to sites instead of links.
Note the absence of the momentum dependence in Eq.~(\ref{eq:Dyson}): The bold Green function remains local on the links in any order of renormalization. It means, in particular, that the only non-zero element for a link between sites $(0,0)$ and $(1,0)$ is $G_{02}$; it can be alternatively denoted as $G_{x}$ and, by $90^{o}$  rotation symmetry of the square lattice, is the same
for all links.
%To treat primed and non-primed Grassmann variables on equal footing, we write the bare Green function in a symmetrized form as $G^{(0)}_1=G^{(0)}_2=G^{(0)} = \sqrt{\zeta}$.

\item The matrix structure of polarization $\Pi$ is similar to that of $V$.
The 0th order expression based on bare Green functions is given by
\begin{equation}
\Pi^{(0)}_{\rm (x,y)}(\alpha, \gamma) = \zeta \begin{bmatrix}
0 & \!\! 0 &  \!\! \delta_{x,1}\delta_{y,0} &  \!\! 0 \\
0 & \!\! 0 &  \!\!  0 &  \!\! \delta_{x,0}\delta_{y,1} \\ \delta_{x,-1}\delta_{y,0} & 0 &0 & 0 \\ 0 & \delta_{x,0}\delta_{y,-1} &  \!\! 0 &  \!\! 0   \end{bmatrix} .
\end{equation}

\item The effective potential $W$ is defined through the Dyson equation
in momentum representation
\begin{equation}
W_{\bm q}(\alpha, \gamma ) = V(\alpha, \gamma ) + \sum_{\mu, \nu} V(\alpha, \mu)
\Pi_{\bm q}(\mu, \nu) W_{\bm q}(\nu, \gamma ) \,. \label{eq:W}
\end{equation}
We expect to see signatures of the ferromagnetic transition in matrix elements of $W_{{\bm q}=0}$
because they directly relate to the divergent uniform susceptibility $\chi$.
\end{enumerate}

\subsection{Zeroth order result}\label{sec:zero}

\begin{figure}[tbp]
\centering
 \includegraphics[width=\columnwidth]{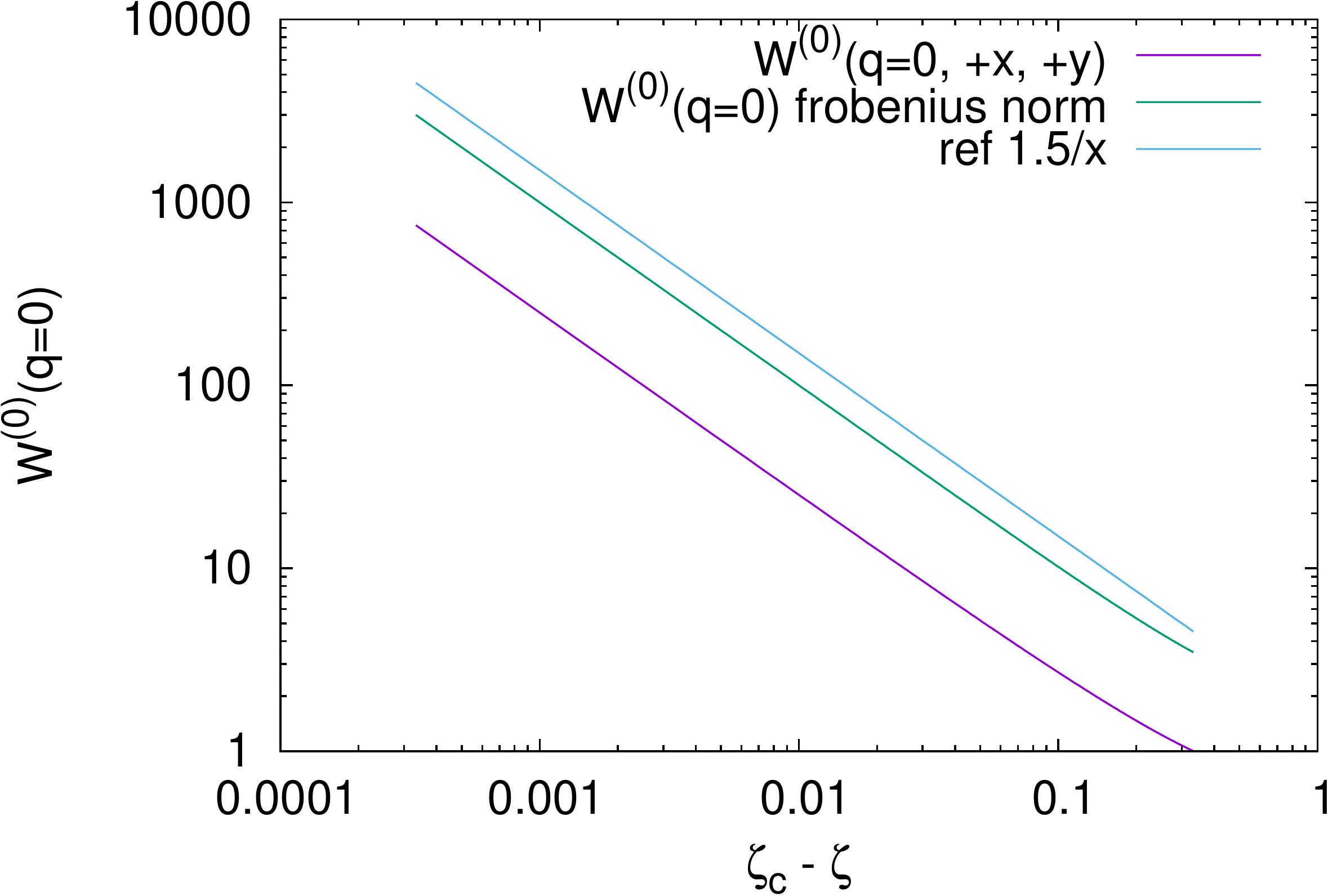}
\caption{Divergence of the 0th order result for $W_{{\bm q}=0}$ at $\zeta_c = 1/3$ is compared
% for two fixed directions (from $+\hat{x}$ to $+\hat{y}$)
with the Frobenius norm and a reference line with power $-1$. 
} \label{fig:order0}
\end{figure}

To obtain the 0th order result, we replace $\Pi$ with $\Pi^{(0)}$ in Eq.~(\ref{eq:W}).
For any $\zeta$ we compute $W_{\bm q=0}$ from Eq.~(\ref{eq:W}) by matrix inversion.
We find a divergence at $\zeta_c = 1/3$ (shown in Fig.~\ref{fig:order0}) that can be also
established analytically. We see that $W$ diverges as $(\zeta_c - \zeta)^{-1}$.
We get the same power law behavior for the $(0,1)$ matrix element as well as for the Frobenius norm---they just differ by a constant factor.
% The same power law dependence is also found for $\beta = {\rm arctanh}(\zeta)$.
It is not surprising that our $\zeta_c$ is below the exact value for this model;
the skeleton approach at 0th order is based exclusively on simple ``bubble''
diagrams in terms of bare Green functions that are all positive, leaving to an overestimate of the critical temperature. Fermionic exchange cycles
and vertexes with negative weights do not contribute at this level of approximation.

\subsection{First order result}\label{sec:one}
We now include the diagrams with one $W$ line for the selfenergy and the polarization.
In real space we find
\begin{eqnarray}
\Sigma^{(1)}_x  & = & \Sigma^{(1)} = - G_x W_{(1,0)}(2,0)= - G W_{(1,0)}(2,0) \nonumber \\
\Pi^{(1)} & = &  G^4 W_{(0,0)}(0,0)  + {\rm cycl.}
\label{eq:2dorder}
\end{eqnarray}
The matrix structure of $\Pi^{(1)}$ is identical to that of $\Pi^{(0)}$ and is not shown here
explicitly. Coupled self-consistent Eqs.~(\ref{eq:2dorder}), (\ref{eq:Dyson}), and (\ref{eq:W}) are
solved by iterations.

\subsection{Second order result} \label{sec:two}

\begin{figure}[tbp]
\centering
 \includegraphics[width=\columnwidth]{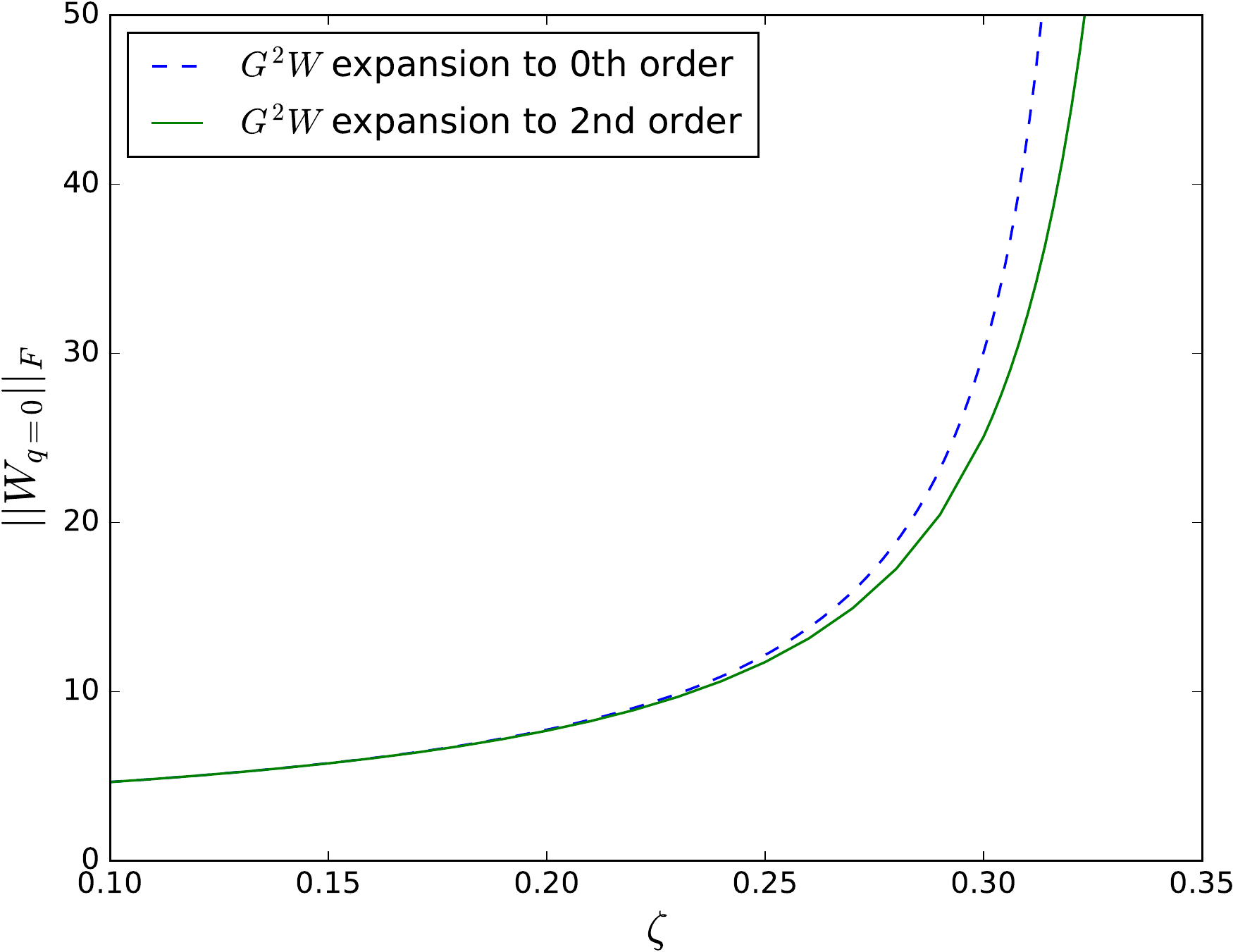}
\caption{(Color online) Shown is the Frobenius norm of $W_{{\bm q}=0}$ (to second order)
on a lattice of size $64 \times 64$.
For comparison, the 0th order result is also shown.
The critical point is found to be at $\zeta_c \approx 0.35$ and the exponent is close to 1.1. } \label{fig:order2}
\end{figure}

As mentioned previously, to account for second-order terms in $\Sigma$,
one goes to the second order graphs for $\Psi$ and removes a $G$ line,
whereas the second-order terms for $\Pi$ are obtained by removing one $W$-line from the
third-order graphs for $\Psi$. The corresponding expressions in real space are
\begin{eqnarray}
\Sigma^{(2)} & = & - W_{(0,0)}(0,0) W_{(0,0)}(2,2) G^3 \nonumber   \\
\Pi_{ (0,0)}^{(2)}(0,0) & = & G^6 W_{(0,0)}(2,2) W_{(1,0)}(0,2) \nonumber \\
\Pi_{ (1,0)}^{(2)}(0,2) & = & G^6 W^2_{(1,0)}(0,2) +  \nonumber \\
{} & {} & G^6 W_{(0,0)}(0, 0) W_{(0,0)}(2,2).
\end{eqnarray}
The remaining non-zero contributions are obtained by invoking discrete lattice symmetries.
Note that to this order the polarization function is extremely local and contains only
same site and n.n. terms. Again, coupled self-consistent GW-equations are solved by
fixed-point iterations. The resulting behavior for $W$ is analyzed in Fig.~\ref{fig:order2}.
The transition point has slightly shifted to larger values of $\zeta$ compared to the zeroth-order result, and the exponent has also slightly increased.

\subsection{Relating ${\Pi}$ to the spin correlation function}

The $G^2W$-expansion scheme treats different bare vertexes (see Fig.~\ref{fig:prediagrams})
on unequal footing: the $V_2$ vertexes are fully dressed, but the $V_4$ vertexes are included perturbatively (we neglected them so far). These higher-rank vertexes have a weight of comparable magnitude to the $V_2$ vertexes (-2 for $V_4$ vs +1 for $V_2$). In addition, the difference in sign between the weights is expected to result in important cancellations between the diagrams and better
convergent series for the spin correlation function (this is how $\zeta_c$ increases towards
its exact value).

Formally, there is no valid reason for neglecting the $V_4$ vertexes altogether. Let us
show how they can be taken care of in the spirit of the shifted action approach.\cite{Rossi2015}
This discussion also gives us the opportunity to explain how the spin correlator is related
to the $G^2W$ skeleton expansion, which is most easily understood in the limit $\zeta \ll 1$.
By assuming that the skeleton sequence (without $V_4$) is solved, we introduce the full
polarization function $\bar{\Pi} (\hat{\alpha}, \hat{\beta} )$ through the Dyson equation
\begin{equation}
\bar{\Pi}_{\bm q}(\alpha, \gamma ) = \Pi_{\bm q}(\alpha, \gamma ) +
\sum_{\mu, \nu} \Pi_{\bm q} (\alpha, \mu) V(\mu, \nu) \bar{\Pi}_{\bm q}(\nu, \gamma ) \,.  \label{eq:Pi_full}
\end{equation}
To be specific, we focus on the n.n. element $\rho_{(1,0)}$; similar manipulations hold
for any other distance. Now consider all diagrams for this correlator without the $V_4$ vertexes
within the $G^2W$ formulation (see Ref.~\onlinecite{Rossi2015}):
\begin{itemize}
\item Put one $V_1$ vertex on the origin site $(0,0)$ and the other $V_1$ vertex on the target site ${\mathbf r}=(1,0)$, see Eq.~(\ref{rhoprim}). There are $4 \times 4 = 16$ different ways of doing that depending on the directions of legs. Connect the legs with $\bar{\Pi}_{\bm r}(\alpha, \gamma)$. For example, in the limit of $\zeta \ll 1$, choosing the ($\alpha\! =\! 0$)-leg on site $(0,0)$ and the $\gamma=2$-leg on site $(1,0)$ results in the contributions $\zeta - 4 \zeta^5 + \ldots $. Similarly, the choice of $\alpha=1$ and $\gamma=1$ leads to the
    contribution $\zeta^3$.
\item Put $V_3$ on $(0,0)$ and $V_1$ on $(1,0)$, and connect all legs with $\bar{\Pi}$ lines.
    There are four ways to orient the $V_3$ vertex and for each one there are two choices for connecting legs with $\bar{\Pi}$ propagators. The leading contribution to $\rho_{(1,0)}$ goes hence as $-8 \zeta^5$.
\item Putting $V_1$ on $(0,0)$ and $V_3$ on $(1,0)$ gives the same contribution by symmetry.
\item Put one $V_3$ vertex on $(0,0)$ and the other $V_3$ vertex on $(1,0)$.
    Now there are 16 ways of orienting both $V_3$ vertexes, and for each orientation there are 15 choices for connecting the legs. These contributions start at order $\propto \zeta^9$.
\end{itemize}

Next, we repeat the above procedure of connecting legs by adding one $V_4$ vertex,
which can be put on any site, after that we can add two $V_4$ vertexes etc. to generate a
perturbative expansion in the number of $V_4$ terms.
Compared to the original bare series in powers of $\zeta$, we have reordered the series:
the effective potential is summing up all $V_2$ vertexes, whereas we expand (and sample in a Monte Carlo framework) in powers of $\lambda_4$.

To illustrate this framework, let us take $\zeta = 0.01$ and recall that in the bare series $\rho_{(1,0)} = \zeta + 2 \zeta^3 + 4 \zeta^5 + 12 \zeta^7 + \ldots$.
The first 3 terms can be reproduced without $V_4$ vertexes and with only 1 $V_3$ on either the origin or the target site, see Figs.~\ref{fig:rho_generic_a}--\ref{fig:rho_generic_d}.
The fifth order coefficient originates from 16 ``simple" diagrams containing just $V_1$ and $V_2$ vertexes without any exchange. The diagrams containing a $V_3$ vertex yield a coefficient $-8$, and the exchange diagrams yield a coefficient $-4$.
On a $16 \times 16$ lattice, the propagators obtained in Sec.~\ref{sec:zero} ({\it i.e.}, to zeroth order) are
\begin{eqnarray}
\bar{\Pi}^{(0)}_{ (x,y)=(1,0) } (0,2) & = & 1.00000002 \times 10 ^{-02} , \\
\bar{\Pi}^{(0)}_{ (x,y)=(1,0) } (1,1) & = & 1.00010011 \times 10 ^{-06} ,\\
\bar{\Pi}^{(0)}_{ (x,y)=(1,0) } (1,2) & = & 1.00080057 \times 10 ^{-10} ,\\
\bar{\Pi}^{(0)}_{ (x,y)=(0,0) } (1,2) & = & 1.00020021 \times 10 ^{-08} .
\end{eqnarray}
We do not mention explicitly other symmetry-related elements.
The sum of all matrix elements for $\bar{\Pi}^{(0)}_{(x,y)=(1,0)}$ is $0.01000200160$. One clearly recognizes the coefficients $1$, $2$ and $16$ for the first-, third- and fifth-order contributions
to the bare series.
Contributions from the $V_3$ vertexes can be estimated from multiplying
$\bar{\Pi}^{(0)}_{ (x,y)=(0,0)}(1,2) \times \bar{\Pi}^{(0)}_{ (x,y)=(1,0)}(0,2)$ which yields $ \approx 10^{-10}$. There are four different diagrams, each with weight $-2$, 
resulting in the above-mentioned
coefficient $-8$.

On a $16 \times 16$ lattice, the propagators obtained in Sec.~\ref{sec:two}  ({\it i.e.}, to second order) are
\begin{eqnarray}
\bar{\Pi}_{ (x,y)=(1,0)}(0, 2) & = & 9.99999980\times 10 ^{-03} \\
\bar{\Pi}_{ (x,y)=(1,0)}(1, 1) & = & 1.00009999\times 10 ^{-06} \\
\bar{\Pi}_{ (x,y)=(1,0)}(1, 2) & = & 1.00120089\times 10 ^{-10} \\
\bar{\Pi}_{ (x,y)=(0,0)}(1, 2) & = & 1.00020005\times 10 ^{-08}
\end{eqnarray}
The sum of all matrix elements for $\bar{\Pi}^{(0)}_{(x,y)=(1,0)}$ is $0.01000200120$. One clearly recognizes the coefficients $1$, $2$ and $12$ for first, third and fifth order contributions to the bare series. For the fifth order contribution, we now obtain $12$ instead of $16$ thanks to the Grassmann exchange contribution that is accounted for properly at this level of approximation.
By adding the $V_3$ diagrams in the way described above we recover the correct result to this
order in $\zeta$ (which is +4).

The first instance of a $V_4$ vertex occurs in order $\zeta^6$ in the bare series. The relevant bare diagrams are the ones for $\rho_{(1,1)}$ with a $V_4$ vertex on site $(1,0)$ (and all cases related by the lattice symmetry). Our bold expansion can correctly account for this contribution if we put a $V_4$ vertex on this site and connect all unpaired legs with $\bar{\Pi}$ propagators.  However, with the propagators obtained in Sec.~\ref{sec:two} we are not supposed to account for all possible diagrams in the bare series to order 6 because our bold expansion in Sec.~\ref{sec:two} is only accurate
up to order $\zeta^3$: Consider again $\rho_{(1,1)}$ and the bare diagrams where exchanges are possible on the links between the sites $(0,0) - (1,0)$ and $(1,0)-(1,1)$. Then there are irreducible non-local contributions that are not accounted for in Sec.~\ref{sec:two} with a positive weight that involves exchanges on both links in a correlated fashion. These contributions would obviously be accounted for in higher order corrections to $\Psi$, when $\Pi$ becomes non-local. This is also seen in the numerics: the $G^2W$ approach to second order yields a coefficient of $6$ for $\zeta^6$ contribution to  $\rho_{(1,1)}$,
% whereas the contribution form $V_3$ and $V_4$ vertexes is $-16-4 = -20$, yielding $6$
which is below the correct value of $10$.

%\subsection{Shifted action approach}\label{sec:shift_action}

%We now wish to include the simplest possible diagram containing a 'cross' (type $V_4$ vertex). There are two different contributions (one is shown in Fig.~\ref{fig:V4}) with a single cross and as little propagator lines as possible,
%%
%\begin{eqnarray}
%\Pi^{(2b)}_{ (x,y)=(0,0)} (\hat{x}, \hat{y}) & = & -2 \Pi_{{\rm full}, (x,y)=(0,0)} (\hat{x}, \hat{y}) \\
%\Pi^{(2b)}_{ (x,y)=(0,0)} (\hat{x}, -\hat{x}) & = & -2 \Pi_{{\rm full}, (x,y)=(0,0)} (\hat{y}, -\hat{y}). \\
%\end{eqnarray}
%%
%The contributions to the irreducible polarization in the other leg-directions follow by lattice symmetry.
%In the selfconsistency loop we now also need to determine $\Pi_{\rm full}$ but for the rest the procedure is the same. The results are shown in Fig.~\ref{fig:order2B}.
%Note that in the bare series the first diagram with 2 $V_4$ vertexes occurs in order 11.

%\begin{figure}[tbp]
%\centering
% \includegraphics[width=0.5\columnwidth]{fig_V4}
%\caption{Simplest diagram with a $V_4$ vertex contributing to the irreducible polarization. Two of its ends are connected via a $\Pi_{\rm full}$ propagator. There are 6 such diagrams. } \label{fig:V4}
%\end{figure}

%\begin{figure}[tbp]
%\centering
% \includegraphics[width=0.4\columnwidth, angle=-90]{W_order2B_size}
%\caption{System size dependence of $W_{{\bm q}=0}$ in order 2B.  } \label{fig:order2B_size}
%\end{figure}

%%%%%%%%%%%%%%%%%%%%%%%%%%%%%%%%%%%%%%%%%%
\section{The Ising model in one dimension}
\label{sec:VIII}

Let us show that the proposed approach solves the 1D Ising model exactly,
both in the bare formulation as well as in the $G^2W$ skeleton formulation.

\subsection{Bare series}
In 1D, the only allowed vertex is $RL$ (the last one in the second line of Fig.~\ref{fig:prediagrams}). It has weight +1. The only allowed endpoints are $L$ and $R$ (the second and fourth vertexes
shown in the first line of Fig.~\ref{fig:prediagrams}). As expected, this means that there are no loops, no fermionic exchanges, and no minus signs in 1D. At order $n$ of the expansion for the spin correlator there is only one contributing diagram with weight $\zeta^n$
(up to the lattice symmetry). The susceptibility is hence
\begin{equation}
T \chi = 1 +2 ( \zeta + \zeta^2 + \ldots) = 1 + 2 \frac{\zeta}{1-\zeta},
\label{eq:1d_bare}
\end{equation}
reproducing the exact solution
with asymptotic behavior $\chi \propto \beta \exp(2 \beta)$ as $T \to 0$.

\subsection{$G^2W$ formulation}

The $G^2W$ skeleton expansion becomes exact already in 0th order,
\begin{eqnarray}
\Pi & = &  \Pi^0 = \zeta \\
\Sigma &= & 0
\end{eqnarray}
which yields $G = G_0 =\sqrt{\zeta}$, $W = V / (1 - V \Pi) = 1/(1-\zeta)$, and also $\Pi = \zeta/(1- \zeta)$. This immediately leads to the same result as in Eq.~(\ref{eq:1d_bare}) when adding the end-point vertexes $L$ and $R$ to $\Pi$.

%%%%%%%%%%%%%%%%%%%%
\section{Conclusion}
\label{sec:IX}

We have developed a general scheme for mapping a broad class of classical statistical
link (plaquette) models onto interacting Grassmann-field theories that can be studied
by taking full advantage of the diagrammatic technique. This mapping, in particular,
would allow to formulate an all-diagrammatic approach to $(d+1)$-dimensional lattice
gauge theories with finite density of fermions. The resulting field-theory looks very
complex because it contains a large number of Grassmann variables with numerous multi-point
interaction vertexes. Moreover, it is generically strongly-coupled at low temperature
meaning that an accurate solution using diagrammatic methods is only possible when
calculations are performed to high order and extrapolated to the infinite-order limit.

The complexity of the problem should not be taken as an indication that the entire idea
is hopeless. Monte Carlo methods were designed to deal with configuration spaces
of overwhelming size and complexity and arbitrary weights. In this sense, diagrammatic Monte Carlo methods simulating the configuration space
of irredicuble connected Feynman graphs are based on the same general principles
and one should not be surprised that they can evaluate the sum of millions of bare
(or skeleton) graphs, enough to attempt an extrapolation to the infinite-order limit.
What makes diagrammatic Monte Calro distinctly unique (apart from working with ever-changing number
of continuous variables without systematic errors) is the radical transformation of
the sign problem. It is completely eliminated in conventional sense because the
thermodynamic limit is taken first. Given that the number of diagrams increases
factorially with their order, finite convergence radius in $\zeta$ is only possible
if same-order diagrams cancel each other to such a degree that at high order
their combined contribution is not increasing factorially. In other words,
non-positive weights are {\it required} for the entire approach to work and we call
it the ``sign-blessing" phenomenon. Diagram weights for Grassmann/fermion fields
alternate in sign depending on the diagram topology; this leads to the sign-blessing
phenomenon for lattice models.

We illustrated the proposed approach by considering the 2D Ising model
as a prototypical example. We have deliberately chosen to work with the
generic formulation to avoid model specific simplifications because our
goal was not to solve the model but to demonstrate how one would proceed
in the general case. The ultimate goal is to explore how this field-theoretical
approach can help with understanding properties of lattice gauge models.

%We live in times of devolution. Soon the Ising model will be 100 years old, an exact solution was found in the 1940s and 70 years later we managed to represent it as a hopelessly complicated strongly-interacting fermionic system. But after enough suffering we could find an approximate solution. Life was made hard where it is easy.
%But this is exactly why we do physics instead using our brains to steal other people money.
%OK, we do waste their money by doing projects like this one but at least we suffer.

\section{Acknowledgement}
\label{sec:X}

We are grateful to A. J. Guttmann for providing us with references to the high-temperature series expansions and U. Wolff for drawing our attention to the work by S. Samuel. This work was supported by the National Science Foundation under the grant PHY-1314735, FP7/Marie-Curie Grant No. 321918 (``FDIAGMC"), and FP7/ERC Starting Grant No. 306897 (``QUSIMGAS").

\bibliographystyle{apsrev4-1}
\bibliography{refs}{}

\end{document}